\documentclass[journal]{IEEEtran}

\usepackage[style=ieee]{biblatex} 
\bibliography{example_bib.bib}    

\usepackage{amsmath}

\usepackage{url}
\usepackage{graphicx}  
\usepackage{float}  
\usepackage{xcolor} 

\usepackage{graphicx}
\usepackage{tabularx}
\usepackage{bm}  
\usepackage[american]{circuitikz}
\usepackage{tikz}
\usetikzlibrary{shapes.geometric, arrows, positioning, calc} 
\usetikzlibrary{shapes, arrows}
\usetikzlibrary{decorations.pathmorphing}
\usepackage{xcolor}
\definecolor{darkblue}{rgb}{0.0, 0.0, 0.5} 

\usepackage{hyperref}
\hypersetup{
    colorlinks=true,
    linkcolor=black,    
    citecolor=darkblue, 
    urlcolor=darkblue   
}
\begin{document}

\title{Characterization of an MPPC-Based Scintillator Telescope and Measurement of Cosmic Muon Angular Distribution}

\author{
    Sahla Manithottathil, 
    Anuj Gupta, 
    Mudit Kumar and 
    Navaneeth Poonthottathil\\ 
    \vspace{1.5ex}
    
    Department of Physics, Indian Institute of Technology Kanpur, India 
}
        

\maketitle
\thispagestyle{empty}
\begin{abstract}
This report presents the design, characterization, and application of a high-sensitivity optical detection system based on plastic scintillators coupled to Multi-Pixel Photon Counters (MPPCs). The primary objective was to evaluate the performance of MPPCs (Silicon Photomultipliers) as robust, low-voltage alternatives to traditional photomultiplier tubes for detecting faint scintillation light. The optoelectronic properties of the sensors were analyzed, including single-photoelectron gain calibration and dark count rate measurements, to optimize the signal-to-noise ratio. By embedding wavelength-shifting fibers to enhance light collection efficiency, the system was configured into a three-fold coincidence telescope. The angular distribution of the cosmic ray muon flux was measured to validate the detector's stability and geometric acceptance. Fitting the experimental data to a $\bm{\cos^n(\theta)}$ distribution yielded an angular exponent of $\bm{n = 1.44 \pm 0.06}$, consistent with literature values. These results demonstrate the efficacy of the MPPC-scintillator coupling for precise photon counting and timing applications in high-energy physics instrumentation.
\end{abstract}


\section{Introduction}

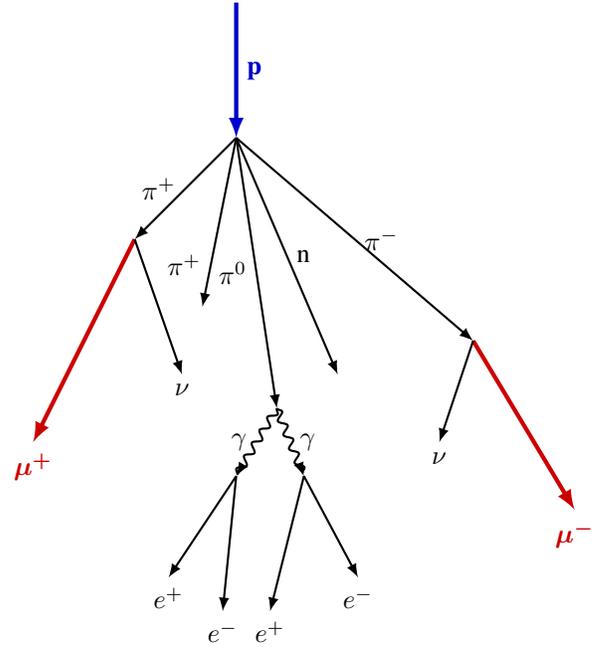
\begin{figure}[h!]
    \centering
    \begin{tikzpicture}[scale=0.9, >=latex]
        \draw[->, ultra thick, blue!80!black] (0, 5) -- (0, 3) node[midway, right] {\textbf{p}};
        
        \coordinate (interaction) at (0,3);

        
        \draw[->, thick] (interaction) -- (-1.5, 1.5) node[midway, left] {$\pi^+$} coordinate (pi_plus1);
        
        \draw[->, thick] (interaction) -- (-0.5, 0.5) node[near end, left] {$\pi^+$};
        
        \draw[->, thick] (interaction) -- (0.6, -1.0) node[midway, left] {$\pi^0$} coordinate (pi_zero);
        
        \draw[->, thick] (interaction) -- (1.5, -0.5) node[midway, right] {n};
        
        \draw[->, thick] (interaction) -- (3.5, 0) node[midway, right] {$\pi^-$} coordinate (pi_minus);

        
        \draw[->, ultra thick, red!80!black] (pi_plus1) -- (-3, -1.5) node[below] {$\bm{\mu^+}$}; 
        \draw[->, thick] (pi_plus1) -- (-0.8, -0.5) node[below] {$\nu$}; 
        
        \draw[->, ultra thick, red!80!black] (pi_minus) -- (5, -2.5) node[below] {$\bm{\mu^-}$}; 
        \draw[->, thick] (pi_minus) -- (3.0, -1.5) node[below] {$\nu$}; 

        \draw[->, thick, decorate, decoration={snake, amplitude=0.5mm, segment length=2mm}] 
            (pi_zero) -- (0, -2) node[midway, left] {$\gamma$} coordinate (gamma1);
            
        \draw[->, thick, decorate, decoration={snake, amplitude=0.5mm, segment length=2mm}] 
            (pi_zero) -- (1.0, -2) node[midway, right] {$\gamma$} coordinate (gamma2);

        
        \draw[->, thick] (gamma1) -- (-1.0, -3.5) node[below] {$e^+$};
        \draw[->, thick] (gamma1) -- (-0.2, -4.0) node[below] {$e^-$};
        
        \draw[->, thick] (gamma2) -- (0.5, -4.0) node[below] {$e^+$};
        \draw[->, thick] (gamma2) -- (1.8, -3.5) node[below] {$e^-$};

    \end{tikzpicture}
    \caption{Schematic representation of a cosmic ray air shower. The primary proton \textbf{p} initiates the cascade. The \textbf{Muons ($\bm{\mu}$)} are highlighted in red as they are the penetrating particles detected in this setup.}
    \label{fig:cascade}
\end{figure}

\IEEEPARstart{E}{arth} is under constant bombardment by cosmic rays, a flux of highly energetic charged particles originating from outer space \nocite{}.
This primary radiation, composed of approximately 89\% protons, 9\% helium nuclei, 1\% electrons and 1\% heavier nuclei, is thought to be accelerated to relativistic speeds by powerful galactic events such as supernovae \cite{yau2008cosmic}.
Upon entering the upper atmosphere, these primary particles collide with air molecules, initiating a cascade of lighter, secondary particles known as an extensive air shower. In these initial interactions, unstable mesons-primarily pions ($\pi$) and kaons ($\kappa$)-are produced.
The charged pions are of particular importance, as they decay almost instantaneously into the particles that are the focus of this study: muons \cite{blanchard2012measurement} refer Fig. \ref{fig:cascade}.

The muon ($\mu$) is a fundamental lepton, similar to an electron but approximately 200 times more massive\cite{venterea2023analysis}. This significant mass is the key to its highly penetrating power. While muons are unstable, with a short intrinsic lifetime of only 2.2 microseconds , those created in the upper atmosphere possess enough energy to travel near the speed of light \cite{le2018cosmic}. Due to the effects of relativistic time dilation, their lifetime in our frame of reference is extended, allowing a significant fraction to survive the journey to the ground \cite{le2018cosmic}. Unlike the electromagnetic and hadronic components of the air shower, which are quickly absorbed by the atmosphere, the muonic component is "penetrating." As a result, the energetic particle flux at sea level is overwhelmingly dominated by these secondary muons, which arrive with a mean energy of about 4 GeV \cite{le2018cosmic}. The directionality of this flux is not uniform; it is known to be most intense from the vertical (zenith) and decreases at larger angles\cite{venterea2023analysis}, a key property that will be investigated in this work.\\

\begin{figure*}[!t]
    \centering
    \includegraphics[trim={0cm 4cm 0cm 3cm}, clip, width=0.8\textwidth]{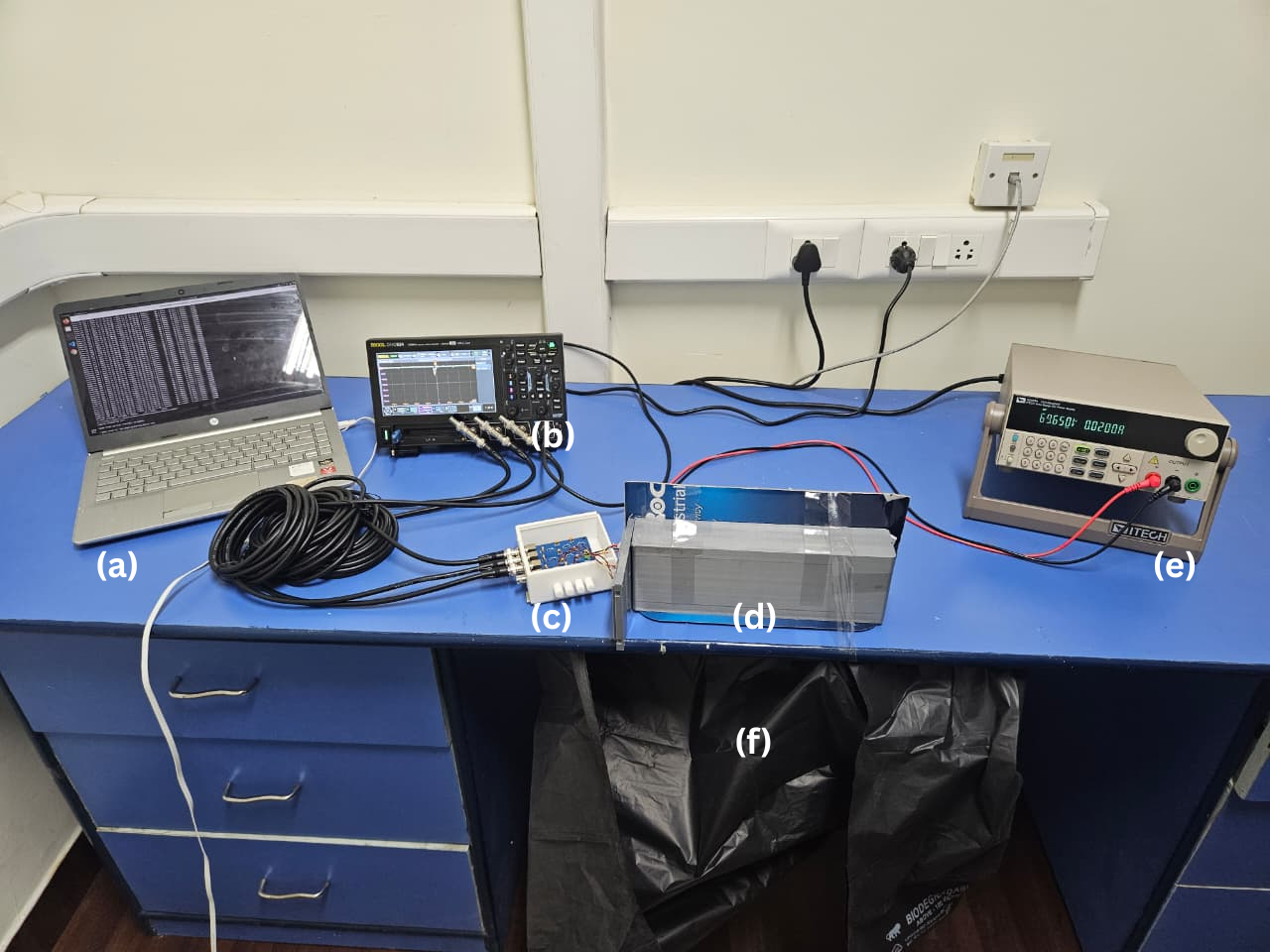}
    
    \caption{The complete experimental setup. \textbf{(a)} Laptop station for data acquisition. \textbf{(b)} RIGOL DHO924 oscilloscope. \textbf{(c)} Interface box for bias voltage distribution and signal readout. \textbf{(d)} The three-fold scintillator telescope assembly (containing the scintillators, WLS fibers, and MPPCs). \textbf{(e)} DC power supply. \textbf{(f)} Light-proof black sheet used to cover the detector assembly during operation to ensure complete darkness.}
    \label{fig:full_setup}
\end{figure*}
   In this experiment, the naturally occurring flux of cosmic muons is used to characterize and validate a prototype muon telescope. The primary objective is to conduct a comprehensive, multi-part investigation into the properties of these particles using a custom-built, three-fold coincidence detector. The first phase of the work involves the complete characterization of the detector system, including optimizing the operating threshold by analyzing the detector's response to background and signal, and experimentally verifying the penetrating power of the detected particles. The main scientific goal is to then use this validated instrument to perform a precise measurement of the muon flux as a function of zenith angle, $\theta$. The resulting angular distribution will be analyzed by performing a comparative fit to several prominent theoretical models, allowing for a quantitative measurement of the physical parameters that govern the muon flux at ground level.\\

\section{Theoretical Framework and Apparatus}
\subsection{Theoretical Principles
}
\subsubsection{Muon Interaction with Matter and Energy Loss}
When high-energy charged particles travel through matter, two primary effects occur: the particle loses energy and it is deflected from its initial direction. If the charged particles are heavy - such as the muons in our case - these effects are primarily due to inelastic collisions with the atomic electrons in the material. Although the energy transfer in a single collision is small, in a dense medium the interaction cross-section is significant. Consequently, many collisions occur per path length, resulting in a substantial cumulative energy loss \cite{le2018cosmic}.

This specific energy loss (stopping power) is described quantum-mechanically by the Bethe-Bloch formula, shown in Equation~\ref{eq:bethebloch} \cite{yau2008cosmic}:

\begin{figure*}[!t] 
    \centering
    \begin{equation}
    -\frac{dE}{dx} = 2\pi N_a r_e^2 m_e c^2 \rho \frac{Z}{A} \frac{z^2}{\beta^2} \left[ \ln \left( \frac{2m_e \gamma^2 v^2 W_{\text{max}}}{I^2} \right) - 2\beta^2 - \delta - \frac{2C}{Z} \right]
    \label{eq:bethebloch}
    \end{equation}
\end{figure*}
\noindent where :\
\begin{itemize}
    \item $N_a$: Avogadro's number
    \item $r_e$: Classical electron radius
    \item $m_e$: Electron mass
    \item $\rho$: Density of the absorbing material (plastic scintillator)
    \item $Z, A$: Atomic number and atomic weight of the material \\
    \item $z$: Charge of the incident particle (muon)
    \item $\beta$: Velocity ratio $v/c$
    \item $\gamma$: Lorentz factor
    \item $W_{\text{max}}$: Maximum energy transfer in a single collision
    \item $I$: Mean excitation potential
    \item $\delta$: Density correction
    \item $C$: Shell correction
\end{itemize}

\begin{figure}[h]
    \centering
    \includegraphics[width=0.5\textwidth]{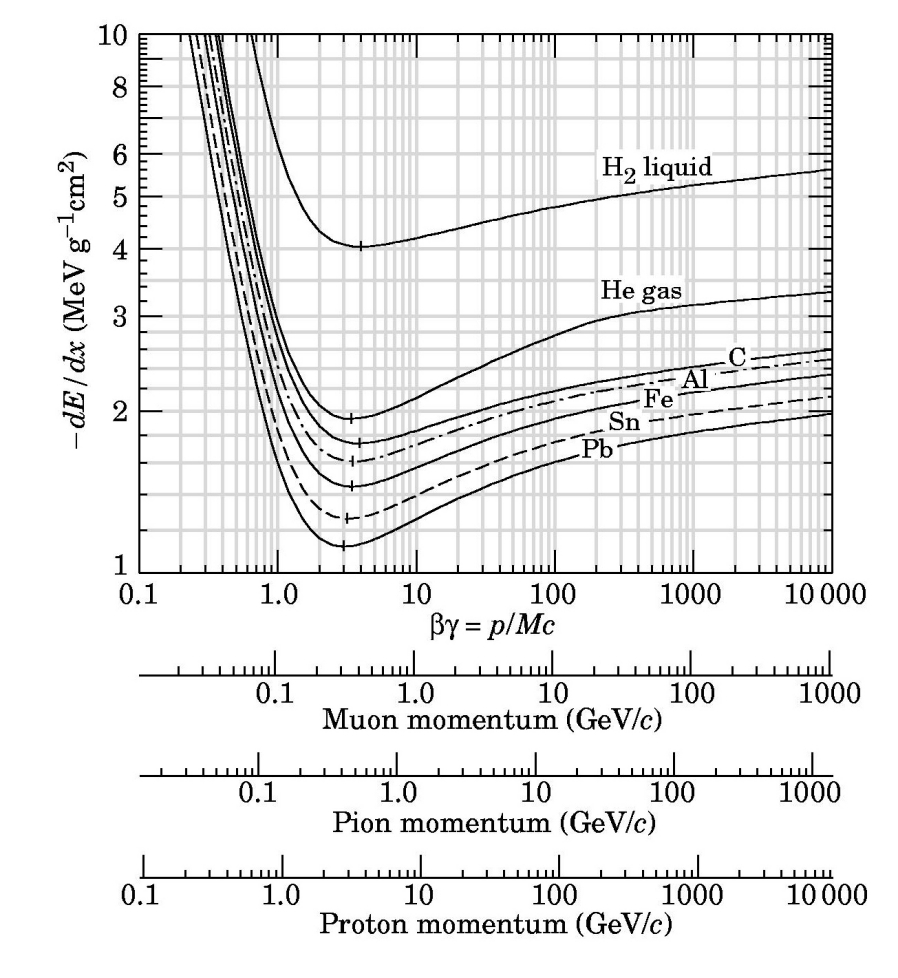} 
    \caption{Stopping power ($dE/dx$) as a function of muon momentum. The curve shows the characteristic minimum, defining the Minimum Ionizing Particle (MIP) region \cite{yau2008cosmic}.}
    \label{fig:bethe}
\end{figure}

When this formula is analyzed as a function of the incident particle's momentum, a distinct pattern emerges, as illustrated in Fig.~\ref{fig:bethe}. The energy loss decreases as energy increases until it reaches a specific minimum point, after which the curve becomes relatively constant. Particles moving with momenta near this minimum are referred to as Minimum Ionizing Particles (MIPs).

Since cosmic ray muons travel at relativistic speeds at ground level, they typically behave as MIPs. This means their energy deposition per unit distance traveled is nearly constant; a common approximation for this value is $2~\text{MeV}/(\text{g}/\text{cm}^2)$ \cite{yau2008cosmic}.

This distinct physical signature is the key to reliably counting cosmic muons while rejecting background noise. Because sea-level muons act as MIPs, they deposit a relatively constant and predictable amount of energy as they pass through the scintillator \cite{leo}. The fundamental principle of scintillation detection states that the amplitude of the output electrical pulse is directly proportional to the energy deposited \cite{le2018cosmic}.

Taken together, these principles form the basis of our triggering strategy. Since muons deposit a consistent, substantial amount of energy, they produce a distinct class of large-amplitude pulses. Conversely, background events (such as detector dark counts) correspond to single-photoelectron events and produce consistently small-amplitude pulses \cite{leo}. This clear separation allows for the effective use of a fixed-level (Leading Edge) discriminator. By performing a threshold scan, we can experimentally identify the boundary between the noise and signal regimes and select an operating threshold that efficiently counts the high-amplitude muon pulses while rejecting the high-rate, low-amplitude background.

\subsubsection{Scintillation Detection with Multi-Pixel Photon Counters (MPPCs)}

In this experiment, plastic scintillators are employed for the detection of muons. As muons travel through the material, they excite the molecules within the plastic base. These excited states are unstable and rapidly de-excite, releasing their excess energy as photons. To ensure efficient detection, the emission spectrum of the scintillator is matched to the spectral sensitivity of the MPPC. Crucially, the quantity of light produced is directly proportional to the energy deposited by the incident particle \cite{leo}. This process occurs on a timescale of nanoseconds, providing the precise timing resolution required for the coincidence technique discussed in the following section.

 The emitted photons are detected by an MPPC, also known as a Silicon Photomultiplier (SiPM), coupled to the scintillator. These devices consist of a dense array of thousands of microscopic Avalanche Photodiodes (APDs) connected in parallel. To detect the faint light signals, the MPPC is operated in ``Geiger mode'', achieved by applying a reverse bias voltage slightly above the breakdown voltage. In this state, a single photon hitting a pixel triggers a massive ``avalanche'' of electrons.

Each pixel within the MPPC acts like a binary switch: if one photon hits, one pixel fires; if one hundred photons hit, one hundred pixels fire simultaneously. The final output current is the sum of all firing pixels, producing a measurable electrical pulse large enough for the oscilloscope to detect. However, thermal agitation can sometimes trigger an avalanche even without incident light, creating ``dark noise'', but these low-amplitude signals are easily filtered out by the trigger threshold. Detailed description of the MPPC operation principles and noise characteristics is provided in ~\cite{hamamatsu_tech_note}.
\subsubsection{Electrical Characteristics of MPPCs
}
To use the MPPC effectively for detecting muons, we need to understand two key properties: how it amplifies the signal (Gain) and how it generates noise (dark count rate).

\begin{figure}[h!]
    \centering
    \includegraphics[width=0.3\textwidth]{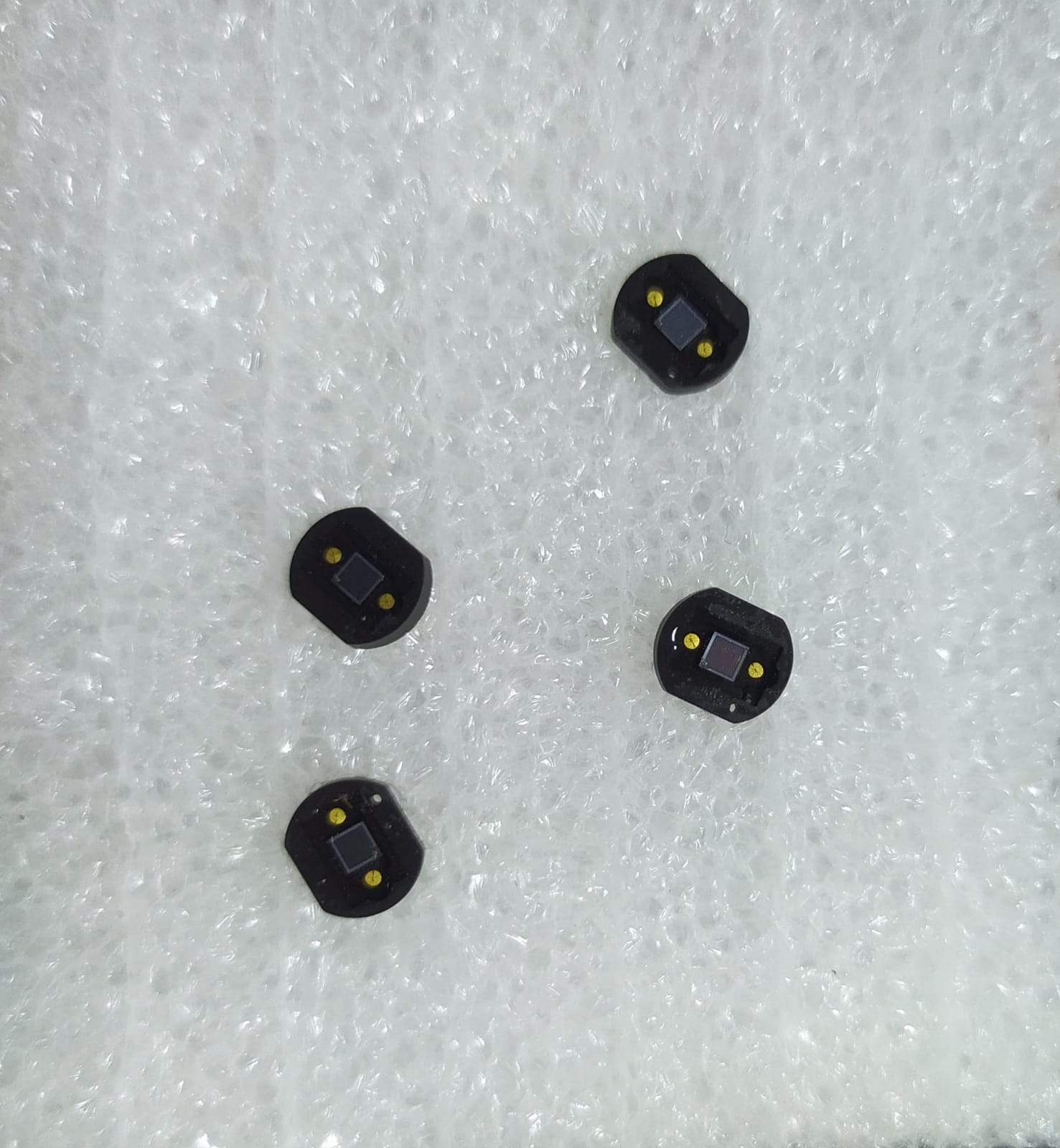}
    \caption{ Hamamatsu MPPC sensors used in the experiment. ~\cite{hamamatsu_s10362} }
    \label{fig:mppc_photo}
\end{figure}

\paragraph{Electric Gain}
The main advantage of the MPPC is its ability to detect extremely weak light signals. It does this by amplifying the signal inside the device, a process known as electric gain. As mentioned in the previous section, this happens through an avalanche effect. Essentially, the gain represents the number of electrons created for every single photon that hits the sensor. This multiplication is what allows us to measure such faint light.

  Mathematically, the gain is defined as the ratio of the total output charge to the elementary charge of a single electron. The formula below shows that the gain depends directly on the voltage we apply to the device \cite{hamamatsu_tech_note}:
\begin{equation}
G = \frac{Q_0}{e} = C_{\text{pixel}} \cdot (V_{\text{operation}} - V_{\text{breakdown}})
\label{eq:sipm_gain}
\end{equation}
In this equation, $V_{\text{operation}}$ is the bias voltage applied to the sensor, and $V_{\text{breakdown}}$ is the threshold voltage necessary to start the avalanche process. The difference between these terms, $(V_{\text{operation}} - V_{\text{breakdown}})$, is commonly referred to as the \textit{overvoltage}. With $C_{\text{pixel}}$ representing the pixel capacitance, the equation shows that the Gain ($G = Q_0/e$) scales linearly with this overvoltage. Therefore, increasing the operation voltage strictly increases the charge output, providing a stronger signal for detection \cite{nguyen2023multi}.
\paragraph{Dark Count Rate (DCR)}
However, this high sensitivity comes with a side effect. Inside the MPPC, pulses are not only created by light. They can also be triggered by heat energy, which knocks electrons loose inside the silicon chip. These false signals are called ``dark pulses'' because they happen even when the detector is in total darkness.

  The problem is that these dark pulses often look exactly like the real signals caused by photons. We cannot tell them apart just by looking at their shape. The number of these false pulses per second is called the Dark Count Rate (DCR).
  Because these pulses come from thermal energy, the rate depends heavily on the temperature of the sensor \cite{streetman}. The formula describing this relationship is given below 
 
\begin{equation}
N_{0.5pe} = A T^{3/2} \exp \left( \frac{-E_g}{2kT} \right)
\label{eq:dark_current_temp}
\end{equation}
In this expression, $T$ represents the absolute temperature in Kelvin, while $k$ is the Boltzmann constant. The parameter $E_g$ denotes the bandgap energy of the semiconductor (approximately 1.12 eV for silicon), and $A$ is a constant related to the pixel geometry and material properties. The presence of the exponential term implies that the dark count rate ($N_{0.5pe}$) is highly sensitive to thermal fluctuations; even a small increase in temperature leads to a significant rise in background noise, highlighting the need for temperature stability during operation.
\subsubsection{The Coincidence Technique for Background Rejection} 
 In the previous section, we discussed setting a trigger threshold to reduce noise. However, relying on a single detector still has limitations, as it can pick up various background signals (e.g., dark noise, natural radiation, electronic interference), creating uncertainty about whether a trigger is a genuine muon or random noise. To overcome this, we arrange multiple detectors in a vertical alignment, creating a ``muon telescope''. This setup operates on an \texttt{AND} logic, recording an event only when all detectors fire simultaneously, a standard technique for signal selection in nuclear physics~\cite{leo}.
   
   \begin{figure}[h!]
    \centering
    \begin{tikzpicture}[scale=0.8, transform shape]
        \tikzstyle{scint} = [draw, fill=cyan!20, minimum width=3cm, minimum height=0.6cm, rectangle]
        \tikzstyle{mppc} = [draw, fill=black!80, minimum width=0.4cm, minimum height=0.4cm, rectangle]
        \tikzstyle{signal} = [draw=blue!80, thick, smooth]

        \node[scint] (s1) at (0, 3) {Top Detector};
        \node[mppc] (m1) at (1.5, 3) {}; 
        
        \node[scint] (s2) at (0, 1.5) {Middle Detector};
        \node[mppc] (m2) at (1.5, 1.5) {}; 
        
        \node[scint] (s3) at (0, 0) {Bottom Detector};
        \node[mppc] (m3) at (1.5, 0) {}; 

        \draw[red, dashed, thick, ->] (-0.5, 4.0) -- (-0.5, -1) node[below] {Muon ($\mu$)};
        
        \node[star, star points=5, fill=yellow, scale=0.5] at (-0.5, 3) {};
        \node[star, star points=5, fill=yellow, scale=0.5] at (-0.5, 1.5) {};
        \node[star, star  points=5, fill=yellow, scale=0.5] at (-0.5, 0) {};

        \node at (4.5, 4.0) {\textbf{Oscilloscope View}};
        
        \draw[->] (2.5, 3) -- (6.0, 3);
        \draw[signal] (2.5, 3.1) -- (3.8, 3.1) -- (3.9, 3.6) -- (4.0, 2.9) -- (4.2, 3.1) -- (6.0, 3.1);
        
        \draw[->] (2.5, 1.5) -- (6.0, 1.5);
        \draw[signal] (2.5, 1.6) -- (3.8, 1.6) -- (3.9, 2.1) -- (4.0, 1.4) -- (4.2, 1.6) -- (6.0, 1.6);
        
        \draw[->] (2.5, 0) -- (6.0, 0);
        \draw[signal] (2.5, 0.1) -- (3.8, 0.1) -- (3.9, 0.6) -- (4.0, -0.1) -- (4.2, 0.1) -- (6.0, 0.1);

        \draw[green!60!black, thick, dashed] (3.9, 3.8) -- (3.9, -0.5) node[below] {Coincidence Window};
        
        \draw[dotted] (m1) -- (2.5, 3.3);
        \draw[dotted] (m2) -- (2.5, 1.8);
        \draw[dotted] (m3) -- (2.5, 0.3);

    \end{tikzpicture}
    \caption{Conceptual diagram of the three-fold coincidence technique. A valid event is recorded only when a single muon passes through all three detectors simultaneously, generating overlapping electrical pulses. This logic effectively filters out random background noise.}
    \label{fig:coincidence_concept}
\end{figure}
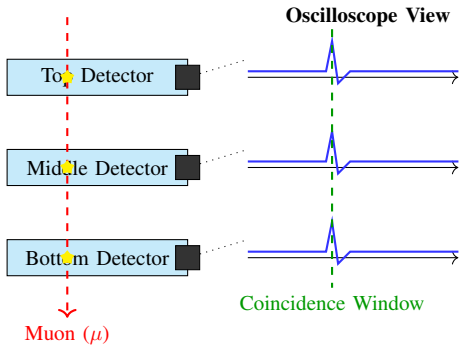
However, since real electronics are not infinitely fast, we must define a short time limit known as the ``coincidence window'' ($\tau$). Occasionally, random noise pulses might overlap within this window just by luck; these are called ``accidental coincidences''. For a three-detector system ~\cite{janossy1944rate}, the rate of these accidents is given by  

\begin{equation}
R_{\text{acc}} \approx 3 \tau^2 R_1 R_2 R_3
\label{eq:accidental_rate}
\end{equation}

\noindent where:
\begin{itemize}
    \itemsep0em
    \item $R_{\text{acc}}$: Rate of accidental coincidences
    \item $\tau$: Coincidence resolving time (time window)
    \item $R_1, R_2, R_3$: Singles counting rates of the three individual detectors
\end{itemize}
Because the coincidence window $\tau$ is extremely small (typically on the order of nanoseconds), the factor $\tau^2$ becomes vanishingly small. Consequently, this makes the rate of accidental coincidences negligible compared to the rate of real cosmic ray muon events.
\subsection{Models for Muon Angular Distribution}
It is well known that the number of muons we detect depends heavily on the direction we are looking. Specifically, the muon count is highest when looking straight up (at the zenith) and drops significantly as we tilt the detector toward the horizon. Many researchers have studied this behavior across different locations, altitudes, and latitudes, finding that while the general pattern is consistent, the exact mathematical description can vary.

In this section, we examine four different models to see which one best describes the data collected in our experiment. We start with the standard $\cos^2\theta$ approximation, which is the most common textbook description of muon flux. We then look at the General $\cos^n\theta$ model used by Pethuraj et al \cite{Pethuraj_2017}, which treats the steepness of the curve as a variable rather than a fixed number.

We also consider a more complex physical model by Shukla and Sankrith\cite{Shukla2016EnergyAA}, which accounts for the curvature of the Earth, and finally, the Schwerdt model \cite{schwerdt_desy}. The Schwerdt model is an empirical formula and includes extra terms to account for background offsets. The following subsections explain the physics and mathematics behind each of these approaches.

\vspace{2mm}
\subsubsection{The $\cos^2(\theta)$ Approximation}

The most common and classical approximation used to describe muon distribution at sea level is the $\cos^2(\theta)$ approximation.

The physical reason behind this approximation is that muons which are produced straight above (at the zenith, $\theta=0$) travel the shortest distance through the atmosphere. However, muons arriving at an angle ($\theta$) must travel through a thicker layer of air. Consequently, the path will be longer for inclined muons; hence, they lose more energy and have a higher chance of decaying before they reach the detector.

This physical effect results in the intensity dropping off as the angle increases, following the formula:

\begin{equation}
    I(\theta) = I_0 \cos^2(\theta)
    \label{eq:cos2_model}
\end{equation}

\noindent where $I(\theta)$ is the intensity at angle $\theta$, and $I_0$ is the intensity at the vertical ($\theta=0$).

However, this model assumes the Earth is flat. Therefore, it works well for small angles but becomes less accurate near the horizon (where $\theta$ approaches $90^\circ$).
\vspace{1.5mm}

\subsubsection{The General $\cos^n(\theta)$ Model (Pethuraj et al. \cite{Pethuraj_2017})}

While $\cos^2(\theta)$ is the standard textbook approximation, it is not a universal law. The exact shape of the distribution can change depending on the energy threshold of the detector and the altitude of the experiment.

To account for these variations, Pethuraj et al. proposed leaving the exponent as a free parameter ($n$) rather than fixing it at exactly 2. The model is expressed as:

\begin{equation}
    I(\theta) = I_0 \cos^n(\theta)
    \label{eq:general_n_model}
\end{equation}

The parameter $n$ describes the steepness of the distribution. We can determine the specific value for $n$ for our detector location by fitting this function to experimental data.

In the study by Pethuraj et al.\ \cite{Pethuraj_2017}, they found a value of $n \approx 2.00 \pm 0.04$ at an altitude of $\sim 160$~m. This confirms that the distribution is indeed close to the theoretical prediction but allows for experimental fine-tuning.
\subsubsection{The Shukla \& Sankrith Model \cite{Shukla2016EnergyAA}}

The problem with the standard $\cos^n(\theta)$ models is that they assume the Earth is a flat plane. Even though this works for angles near the zenith, as we get close to the horizon ($\theta > 70^\circ$), this assumption fails because it implies the atmosphere becomes infinitely thick at $90^\circ$ (which would mean zero muons could ever reach us).

Shukla and Sankrith derived a more accurate formula that accounts for the curvature of the Earth. Instead of a simple flat layer, they model the atmosphere as a spherical shell. Their model replaces the simple cosine term with a path-length factor $D(\theta)$. The formula is:

\begin{equation}
    I(\theta) = I_0 D(\theta)^{-(n-1)}
    \label{eq:shukla_model}
\end{equation}

\noindent with

\begin{equation}
    D(\theta) = \sqrt{\frac{R^2}{d^2}\cos^2\theta + 2\frac{R}{d} + 1} - \frac{R}{d}\cos\theta
    \label{eq:shukla_D}
\end{equation}

Here, $D(\theta)$ represents the effective path length through the curved atmosphere, $R$ is the radius of the Earth, and $d$ is the height of the atmosphere production layer. This allows the model to remain valid even at very large zenith angles (near the horizon), where the standard cosine models break down.

In their paper, they found the exponent parameter to be approximately $n \approx 3.09 \pm 0.03$. (Note that $n$ here relates to the power index of the energy spectrum).
\subsubsection{The Schwerdt Model \cite{schwerdt_desy}}

Unlike the previous models, which are based on the geometry of the atmosphere, this model is empirical. This means it is designed primarily to fit the observed data points mathematically, rather than deriving the curve from physics principles alone.

Schwerdt proposed modifying the standard cosine-squared law by introducing four free parameters to allow for adjustments in amplitude, frequency, phase, and offset. The model is expressed as:

\begin{equation}
    I(\theta) = a \cos^2(b\theta + c) + d
    \label{eq:schwerdt_model}
\end{equation}

with $a$ representing a vertical stretch, $b$ a horizontal stretch,
$c$ a horizontal stretch, and $d$ a vertical shift. \cite{schwerdt_desy}

The most significant addition is the parameter $d$. In standard physical models, the flux at $90^\circ$ (the horizon) is predicted to be exactly zero. However, in real experiments, we often detect a small non-zero count at the horizon due to background radiation from the ground or scattered particles. The $d$ term allows the model to account for this constant background noise, often resulting in a much better statistical fit to the experimental data than the rigid theoretical models.
\subsection{Apparatus}

The complete experimental apparatus used for this study is shown in Fig.~\ref{fig:full_setup}. The system integrates the scintillation detectors, signal processing electronics, and data acquisition hardware into a compact tabletop assembly. The specific components and their configuration are detailed in the following subsections.
\subsubsection{Detector Components: Scintillators and MPPCs}
\paragraph{Plastic Scintillators and Wavelength Shifters}

For the detection of muons in this experiment, we relied on organic plastic scintillators, which are one of the most common and robust tools used in high-energy physics \cite{leo}. We chose plastic material because it is durable, chemically stable, and easy to manufacture.
\begin{figure}[h!]
    \centering
    \includegraphics[width=0.7\linewidth]{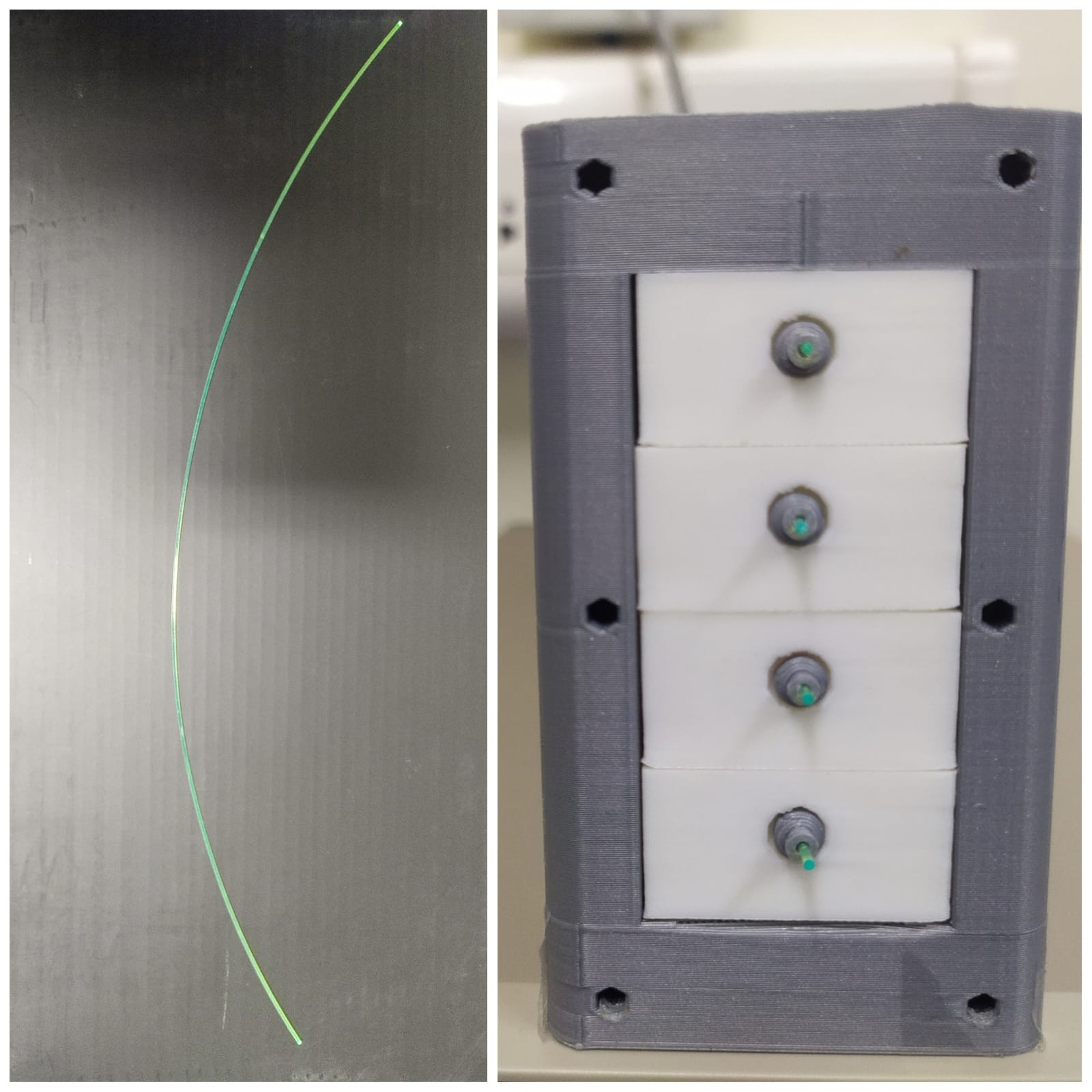}
    
    \caption{(Left) The wavelength-shifting (WLS) fiber used to collect scintillation light. (Right) The assembled detector stack showing the four white plastic scintillator blocks mounted in the 3D-printed frame. The green tips of the fibers are visible at the center of each block, ready to be coupled to the MPPCs.}
    \label{fig:fiber_stack}
\end{figure}
In our specific setup, we employed four rectangular plastic bars, each measuring 25~cm in length, 2.5~cm in width, and 1.28~cm in thickness.

When a cosmic muon passes through one of these bars, it interacts with the plastic to produce a faint flash of blue light \cite{leo}. However, since our light sensor is very small compared to the size of the bar, simply attaching it to the plastic would result in a significant loss of signal. To capture this light efficiently, we embedded a wavelength-shifting fiber (see Fig. \ref{fig:fiber_stack}), about 1~mm in diameter, along the center of each scintillator. This fiber performs a critical function: it absorbs the original blue light and re-emits it as green light. This green color is actually visible to the naked eye at the tips of the fiber. By shifting the color, the fiber acts as a light guide, trapping the signal inside and funneling it directly to the end of the bar where our sensor is mounted \cite{leo}.

\paragraph{Multi-Pixel Photon Counters (MPPCs)}

The device used to read the optical signal is the Hamamatsu S10362-11 Series Multi-Pixel Photon Counter (MPPC) \cite{hamamatsu_s10362}. This device acts as the modern equivalent of a Photomultiplier Tube (PMT). While traditional PMTs are bulky, fragile, and require high voltages, this MPPC is a compact silicon device that is physically robust and operates at a relatively low bias voltage (approximately 70~V).

The specific sensor employed in our experiment has a tiny effective active area of $1~\text{mm} \times 1~\text{mm}$. Despite its small size, it contains a dense array of avalanche photodiodes (pixels). Depending on the specific model variation used, the sensor contains between 100 and 1600 individual pixels. These pixels operate in Geiger mode, effectively acting as binary switches that fire when struck by a photon.

According to the manufacturer specifications \cite{hamamatsu_s10362}, the peak sensitivity of this sensor is at 440~nm. However, its spectral response range is broad (320 to 900~nm), which allows it to efficiently detect the green light emitted by the wavelength-shifting fiber \cite{hamamatsu_tech_note}. This combination of spectral matching and high internal gain (ranging from $10^5$ to $10^6$) ensures that even the faint light pulses from a single muon are amplified enough to be recorded by our electronics.
\paragraph{Biasing and Signal Readout Circuit}

\begin{figure}[h!]
    \centering
    \includegraphics[angle=-90, width=0.3\textwidth]{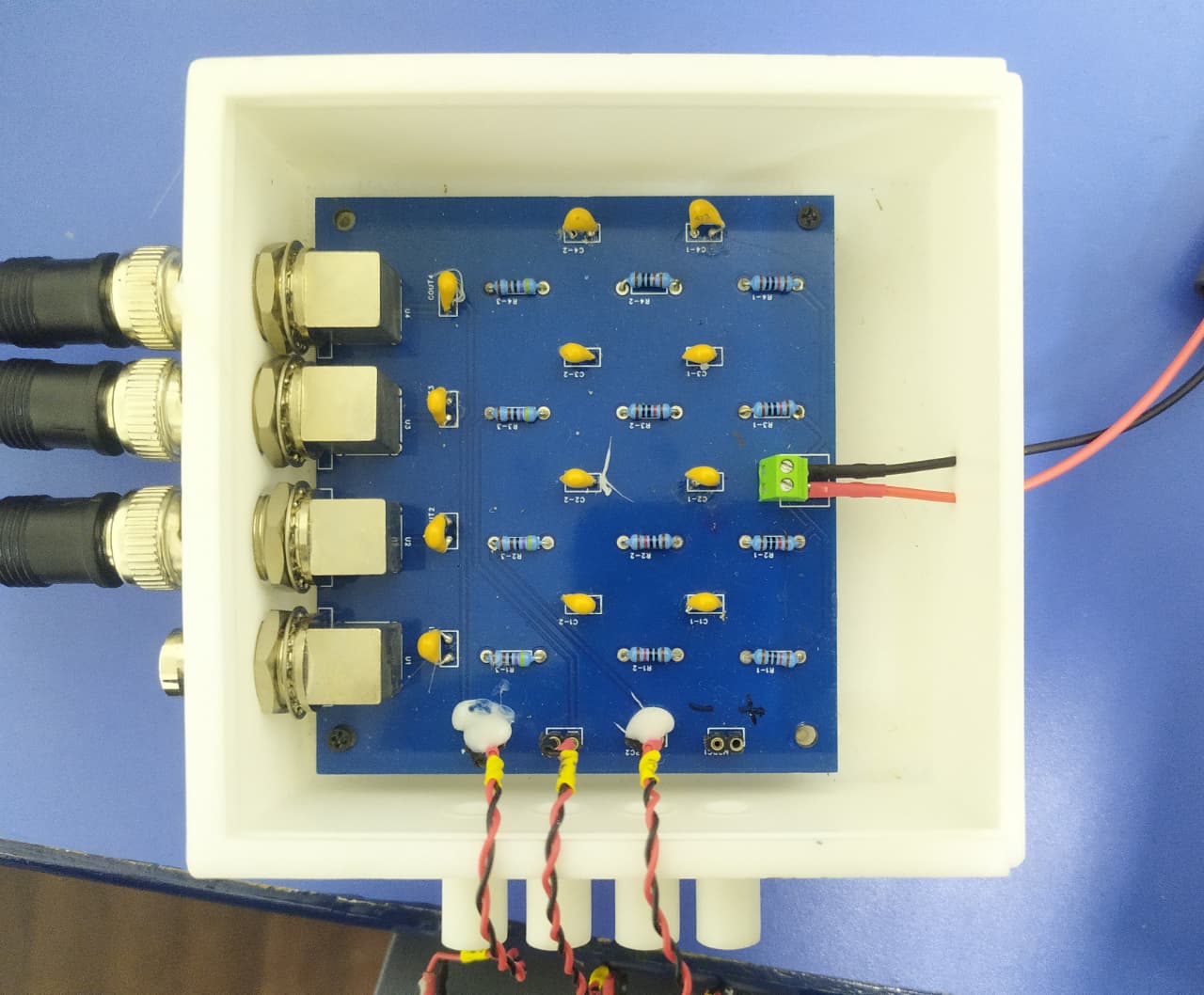}
 
    \vspace{0.5cm} 
    
    \begin{circuitikz}[scale=0.9, transform shape]
        \draw (-3.5, 6) 
            to[pD, l_=MPPC, invert] (-3.5, 4) 
            node[ground]{};
        
        \draw (-3.5, 6) -- (0,6); 

        \draw (0,0) node[below] {$V_{DC}$} 
            to[short, o-] (0,0.5)
            to[R, l=$10\,\text{k}\Omega$] (0,2.5) coordinate(node1)
            to[R, l=$10\,\text{k}\Omega$] (0,4.5) coordinate(node2)
            to[R, l=$4.7\,\text{k}\Omega$] (0,6); 

        \draw (node1) to[C, l={$C_3=47\,\text{nF}$}] (3, 2.5) coordinate(right1);
        
        \draw (node2) to[C, l={$C_2=47\,\text{nF}$}] (3, 4.5) coordinate(right2);

        \draw (right2) -- (right1) -- (3, 0.5) node[ground]{};

        \draw (0,6) to[C, l={$C_1=100\,\text{nF}$}] (3,6)
            to[short, -o] (4,6) node[right] {Out};

    \end{circuitikz}
    
    \caption{(Top) The custom circuit board used to power the four detectors. (Bottom) Circuit schematic of the bias-tee filter for a \textbf{single channel}. The actual PCB implements four identical copies of this circuit in parallel. It uses a multi-stage RC network ($10\,\text{k}\Omega, 47\,\text{nF}$) to filter power supply noise and a $100\,\text{nF}$ capacitor to couple the signal pulse to the output.}
    \label{fig:bias_circuit}
\end{figure}

To operate the MPPCs, a custom circuit board was designed to apply the necessary bias voltage and extract the signal. As shown in Fig.~\ref{fig:bias_circuit}, the circuit uses a network of resistors and capacitors to connect the power supply to the sensor.

The main function of this circuit is to provide a stable DC voltage to the MPPC while allowing the fast signal pulses (generated by muons) to pass through to the output. The capacitors also help to reduce electronic noise from the power supply, ensuring a clean signal for the oscilloscope.

\subsubsection{Data Acquisition System and Triggering Logic}

To truly understand the physical events occurring inside the scintillators, we need a way to visualize the electrical signals in real time. This is the fundamental role of the oscilloscope. While these instruments are typically found in basic electronics labs for studying simple, repeating waves, in our experiment they serve a much more advanced purpose. They act as the primary window into the detector, capturing the ultra-fast and random pulses generated when subatomic particles interact with the plastic.

In the past, analyzing these types of nuclear signals required racks of specialized and expensive digitizing hardware. However, digital technology has evolved to the point where a single, high-performance desktop oscilloscope can now replace those bulky devices \cite{nguyen2023multi}. Modern scopes offer advanced features that allow them to act as both the eyes of the experiment, visualizing the pulses, and the brain, performing the logic required to identify valid data.

\begin{figure}[h!]
    \centering
    \includegraphics[width=0.45\textwidth]{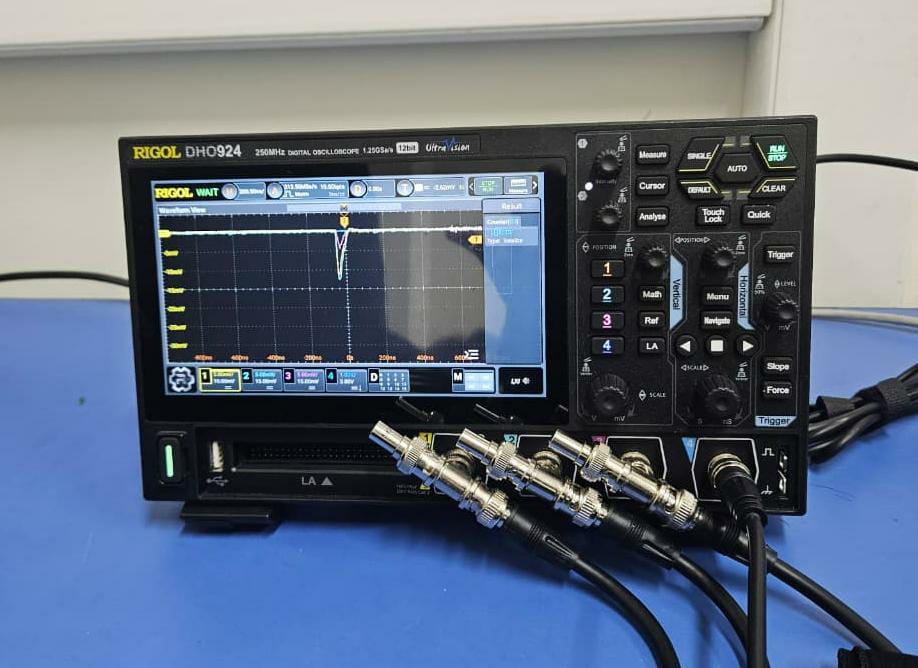}
    \caption{The RIGOL DHO924 High-Resolution Oscilloscope used in the experiment. Its 12-bit vertical resolution allows for precise characterization of the small amplitude MPPC pulses and implementation of the coincidence logic.}
    \label{fig:scope}
\end{figure}

For this specific setup, we selected the RIGOL DHO924  (see Fig. \ref{fig:scope}). This instrument represents a significant improvement over standard laboratory equipment because it is a high-resolution 12-bit oscilloscope. Most common scopes only offer 8-bit resolution, which can make small signals look grainy or pixelated. The 12-bit resolution of our device provides 16 times more vertical precision, allowing us to see the signal with exceptional clarity. This precision allows us to set a distinct voltage threshold to reject low-amplitude thermal noise, ensuring we only analyze valid signals. Technical specifiations are given at Table \ref{tab:rigol_specs}.

The device features four analog channels with a bandwidth of 250~MHz, which is perfectly matched to our multi-layer detector setup. It samples the signal at a rate of 1.25~GSa/s, or 1.25 billion samples per second, ensuring that the sharp, nanosecond-scale rising edge of the MPPC pulse is captured accurately. Furthermore, we utilize the oscilloscope's advanced internal triggering to perform the coincidence logic described in our theory section. Instead of recording every tiny blip, we instruct the scope to only save data when the signals from all active channels overlap in time, effectively filtering out the noise and keeping only the physics events we want to study.
\begin{table}[h!]
    \centering
    \caption{Technical specifications of the RIGOL DHO924 Oscilloscope.}
    \label{tab:rigol_specs}
    \renewcommand{\arraystretch}{1.3} 
    \begin{tabularx}{\linewidth}{@{} p{3.5cm} X @{}} 
        \hline
        \textbf{Feature} & \textbf{Specification (Rigol DHO924)} \\
        \hline
        Bandwidth & 250~MHz \\
        Sampling rate & 1.25~GSa/s (Max Real-time) \\
        Waveform capture rate & 1,000,000~wfms/s (UltraAcquire Mode) \\
        Number of measurement channels & 4 Analog Channels + 16 Digital Channels \\
        Memory depth & 50~Mpts (Max) \\
        Trigger type & Edge, Slope, Pulse Width, Window, Runt, Timeout, Pattern, Video, Duration, Delay, Setup/Hold, Nth Edge \\
        Serial Trigger \& Decoder & I2C, SPI, RS232/UART, CAN, LIN \\
        Digital channel & 16 Channels (Function is Standard; PLA2216 Probe is Optional) \\
        Clock generation module (AFG) & No (Available on DHO924\textbf{S} model only) \\
        Bode Plot & No (Available on DHO924\textbf{S} model only) \\
        Connectivity / Adaptors & LAN, USB Host, USB Device, HDMI, AUX OUT \\
        Display & 7-inch capacitive multi-touch screen ($1024 \times 600$ pixels) \\
        Weight & Approx. 1.78~kg \\
        \hline
    \end{tabularx}
\end{table}
\subsubsection{The Three-Fold Coincidence Telescope Geometry}

To measure the direction of incoming muons, we arranged the three scintillator paddles into a telescope configuration. In this setup, the detectors are stacked vertically and aligned parallel to one another, with the top, middle, and bottom counters placed directly in line. This alignment defines a specific ``line of sight'' for the experiment. For a coincidence event to be recorded, a muon must travel in a straight trajectory through all three layers. This requirement effectively filters out background radiation particles that arrive from random directions.

The detectors are secured to a rigid mechanical frame that maintains a fixed distance between the layers. Crucially, this frame is designed to rotate around a horizontal axis, allowing us to point the telescope toward different sections of the sky. For this experiment, we collected data at four specific zenith angles: $0^\circ$ (Vertical), $30^\circ$, $60^\circ$, and $90^\circ$ (Horizontal). To ensure accuracy, the angle of the frame was verified using a digital inclinometer or protractor before each data run.

The geometry of the stack-specifically the physical size of the detectors ($25~\text{cm} \times 2.5~\text{cm}$) and the vertical distance between the top and bottom counters-determines the solid angle of acceptance. This geometry essentially restricts the detector's field of view. A tall telescope (large separation) sees a small patch of sky, while a compact telescope (small separation) observes a wider area. Since the frame is rigid, this geometric factor remains constant for all angular measurements, ensuring that the comparison between the $0^\circ$ flux and the $90^\circ$ flux is scientifically valid.
\section{Detector Characterization and Performance}

\subsection{MPPC Signal Characterization}

The characterization was performed without an external light source. Instead, the MPPC's own thermally generated dark counts were used as a reliable, built-in signal source. These dark counts are not just random noise; they produce discrete, quantized pulses with amplitudes corresponding to 1~photoelectron (p.e.), 2~p.e., and so on, behaving similarly to light-induced signals~\cite{hamamatsu_tech_note}.

\subsubsection{Single Photoelectron (p.e.) Identification and Gain Calculation}

The most important first step in signal detection is to find the fundamental ``unit'' of the signal. By setting the oscilloscope to trigger on the smallest possible signals from the detector in complete darkness, the characteristic 1~p.e.\ dark count pulse was identified and recorded. This is the key calibration constant derived from the system's own noise.

\begin{figure}[h!]
    \centering
    
    \includegraphics[width=0.5\textwidth]{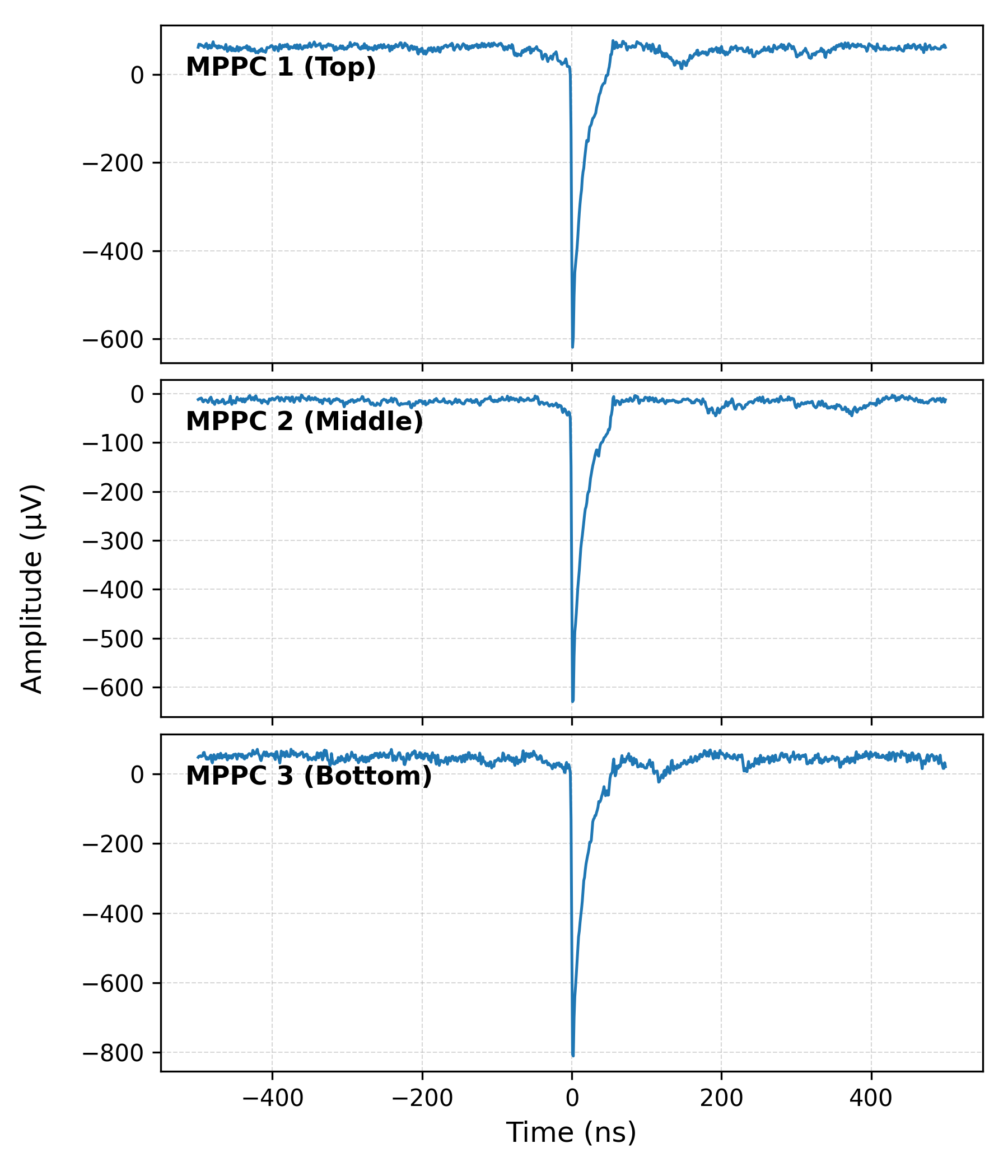}
    
    \caption{Typical single photoelectron (1~p.e.) waveforms for the three detector channels. The pulses exhibit a characteristic amplitude in the range of $580\text{--}660~\mu\text{V}$ with a fast rise time and exponential decay. (Note: Data for MPPC 1 and MPPC 2 were normalized to account for optical crosstalk events captured during the specific acquisition window, ensuring the fundamental 1~p.e.\ pulse shape is represented).}
    \label{fig:1pe}
\end{figure}

The total charge for each 1~p.e.\ event was calculated by integrating the area under the measured voltage pulse. The electrical gain was then determined by dividing this charge by the elementary charge of an electron. The results for each channel are summarized in Table \ref{tab:gain_results}.

\begin{table}[h!]
    \centering
    \caption{The single photoelectron charge and calculated electrical gain for each MPPC channel. (Note: Values for Channels 1 and 2 were derived by normalizing 2~p.e.\ crosstalk events).}
    \label{tab:gain_results}
    \renewcommand{\arraystretch}{1.3} 
    \begin{tabular}{ccc} 
        \hline
        \textbf{Detector ID} & \textbf{1 p.e. Charge (C)} & \textbf{Gain} \\
        \hline
        Channel 1 & $2.89 \times 10^{-13}$ & $1.81 \times 10^6$ \\
        Channel 2 & $2.79 \times 10^{-13}$ & $1.74 \times 10^6$ \\
        Channel 3 & $4.87 \times 10^{-13}$ & $3.04 \times 10^6$ \\
        \hline
    \end{tabular}
\end{table}

The measured electrical gain was found to be on the order of $10^6$, consistent with the typical performance of these devices \cite{hamamatsu_tech_note}. This measured 1~p.e.\ dark pulse voltage was then used as the basis for defining all other threshold levels.

\subsubsection{Dark Count Rate (DCR) Measurement}

With the detector in darkness, the trigger threshold was set to the standard 0.5~p.e.\ level (half the voltage of the 1~p.e.\ dark pulse)\cite{hamamatsu_tech_note}. The total dark count rate was measured and the results for each channel are shown in Table \ref{tab:dcr_rates}.

\begin{table}[h!]
    \centering
    \caption{Measured Dark Count Rate (DCR) for each MPPC at a 0.5 p.e. threshold.}
    \label{tab:dcr_rates}
    \renewcommand{\arraystretch}{1.3}
    \begin{tabular}{cc} 
        \hline
        \textbf{Detector ID} & \textbf{Dark Count Rate (DCR) (kHz)} \\
        \hline
        MPPC 1 & 51 \\
        MPPC 2 & 56 \\
        MPPC 3 & 75 \\
        \hline
    \end{tabular}
\end{table}

The total dark count rate was measured to be in the tens of kHz range. This high rate represents the total flux of single-electron noise events and serves as the baseline rates ($R_1, R_2, R_3$) for calculating the theoretical accidental coincidence rate.
\subsection{Threshold Optimization and Selection}

For the main experiment, finding the optimal trigger setting was crucial to effectively ignore random background noise without missing genuine muon signals. We analyzed the individual detector noise rates and the final coincidence rate. The noise rates of single detectors, working independently at various thresholds, are summarized in Table~\ref{tab:singles_rate}.

\begin{table}[h!]
    \centering
    \caption{Measured singles rate (in kHz) for each detector at various thresholds.}
    \label{tab:singles_rate}
    \renewcommand{\arraystretch}{1.3}
    \begin{tabular}{cccc} 
        \hline
        \textbf{Threshold (p.e.)} & \textbf{MPPC 1} & \textbf{MPPC 2} & \textbf{MPPC 3} \\
        \hline
        1 & 35 & 27 & 35 \\
        2 & 6.5 & 3.8 & 5.86 \\
        3 & 1.2 & 0.5 & 0.916 \\
        4 & 0.24 & 0.08 & 0.14 \\
        \hline
    \end{tabular}
\end{table}

As the Table \ref{tab:singles_rate} shows, the singles rates are very high at the 1 and 2~p.e.\ levels. These high rates of random noise are the main reason we would get false ``accidental'' coincidences.

To ensure our data was free from background noise, we calculated the theoretical rate of accidental coincidences. This calculation relies on the singles rates of the individual detectors and the coincidence window of the oscilloscope ($\tau = $ 200~ns).

Based on the data in Table \ref{tab:singles_rate}, the singles rates at the 3~p.e.\ threshold are approximately $R_1 = 1.2$~kHz, $R_2 = 0.5$~kHz, and $R_3 = 0.92$~kHz. For our main three-fold telescope configuration, the predicted accidental rate is:

\begin{equation}
    R_{\text{acc}} = 3 \tau^2 R_1 R_2 R_3 \approx 6.6 \times 10^{-5}~\text{Hz}
\end{equation}

This value is vanishingly small (effectively zero). It proves that by using a three-fold coincidence, we can safely operate at the lower 3~p.e. threshold. This setting allows us to maximize our detection efficiency for real muons while keeping accidental noise completely negligible. Therefore, 3~p.e.\ was selected as the operating threshold for the angular distribution measurements.

\begin{figure*}[!t] 
    \centering
    \includegraphics[width=0.85\textwidth]{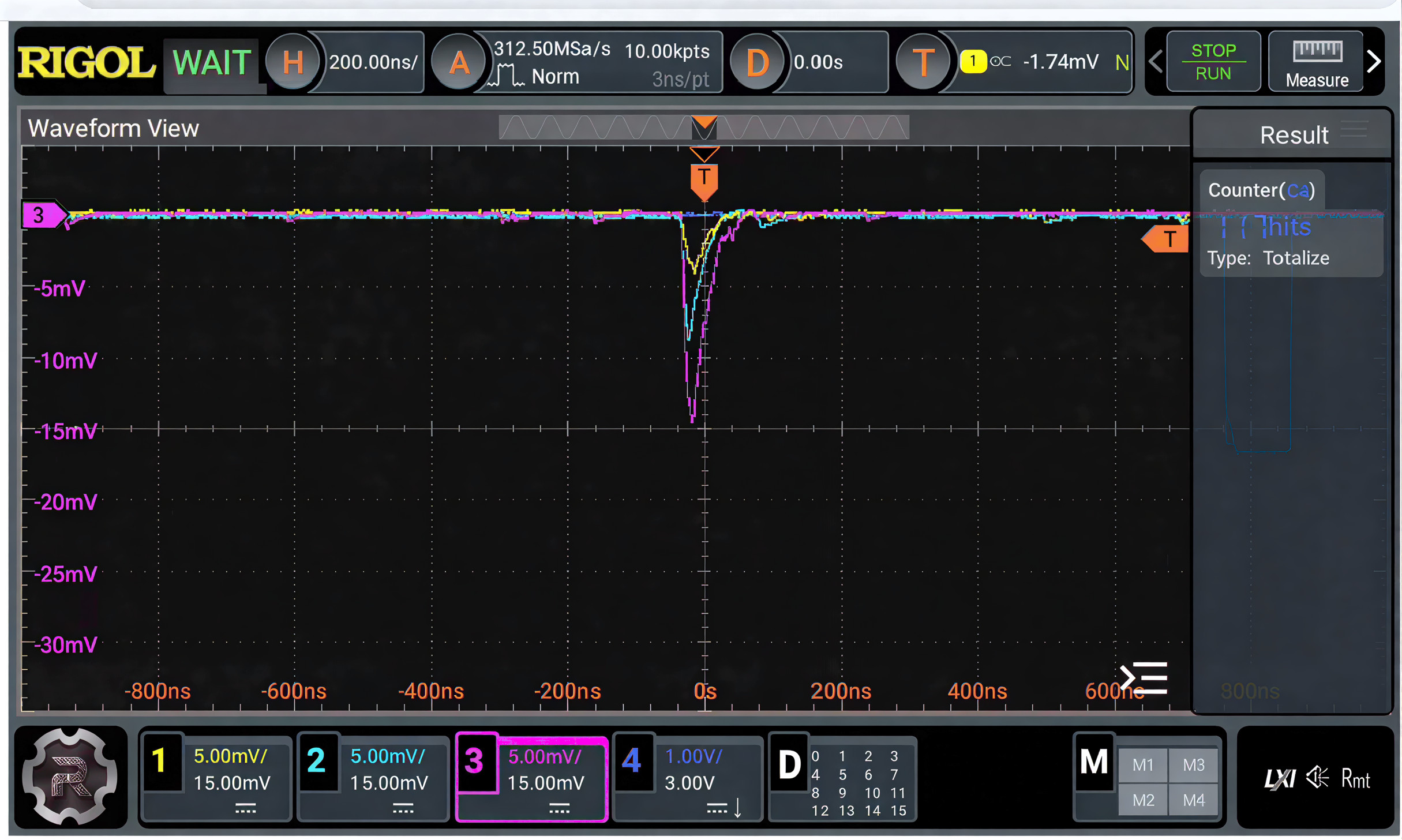}
    
    \caption{Oscilloscope screen capture showing a simultaneous trigger event on three channels (Yellow: Ch1, Cyan: Ch2, Purple: Ch3). This illustrates the coincidence logic: a valid event is recorded only when the electrical pulses from all three detectors overlap in time.}
    \label{fig:scope_trace}
\end{figure*}

Next, we measured the actual two-fold and three-fold coincidence rates, with the results shown in Table \ref{tab:coincidence_freq}. A representative waveform showing the simultaneous firing of three detector channels as observed on the oscilloscope screen is shown in Fig. \ref{fig:scope_trace}.

\begin{table}[h!]
    \centering
    \caption{Measured coincidence frequencies for different photoelectron thresholds.}
    \label{tab:coincidence_freq}
    \renewcommand{\arraystretch}{1.4}
    \begin{tabular}{|l|c|c|c|}
        \hline
        \multicolumn{4}{|c|}{\textbf{Frequency of Coincidence}} \\
        \hline
        \textbf{Threshold} & \textbf{1 p.e.} & \textbf{2 p.e.} & \textbf{3 p.e.} \\
        \hline
        MPPC 1 \& MPPC 3 & 6.5 Hz & 0.05 Hz & No coincidence \\
        \hline
        MPPC 1, 2 \& 3 & 0.166 Hz & No coincidence & No coincidence \\
        \hline
    \end{tabular}
\end{table}

When we look at the results in Table~\ref{tab:coincidence_freq}, we see a dramatic drop in coincidence rates as the threshold increases. Crucially, at the 3~p.e.\ level, the accidental coincidences caused by dark noise essentially disappear. This indicates that any event triggering the detector at this level is almost certainly a real particle, rather than random electronic noise.

For this reason, we chose 3~p.e.\ as our standard operating threshold. It represents the ideal balance point: high enough to silence the background noise, but low enough to ensure we capture the vast majority of genuine muon signals during the main experiment.

\subsection{Verification of Muon Penetrating Power}

To validate that our signal consists of highly penetrating particles (muons) and to select the best final trigger configuration, we compared the rates from two different two-fold coincidence setups~\cite{le2018cosmic}. These two configurations were:
\begin{itemize}
    \item ``Surface Trigger'' (MPPC 1 \& MPPC 2): This setup is sensitive to any particle that can penetrate the first scintillator.
    \item ``Penetrating Trigger'' (MPPC 1 \& MPPC 3): This requires a particle to pass through the entire four-scintillator stack.
\end{itemize}

The results are summarized in Table \ref{tab:penetration_rates}.

\begin{table}[h!]
    \centering
    \caption{COMPARISON OF MEASURED RATES.}
    \label{tab:penetration_rates}
    \renewcommand{\arraystretch}{1.3}
    \begin{tabular}{l p{4cm} c} 
        \hline
        \textbf{Trigger} & \textbf{Description} & \textbf{Rate (Hz)} \\
        \hline
        Surface & Coincidence between top two scintillators & $0.343 \pm 0.010$ \\
        Penetrating & Coincidence between top and bottom & $0.189 \pm 0.007$ \\
        \hline
    \end{tabular}
\end{table}

The ``Surface'' trigger (S1 \& S2) measured a significantly higher rate than the ``Penetrating'' trigger (S1 \& S4). This is expected. The surface trigger detects both penetrating muons and the non-penetrating ``soft'' component of cosmic rays (likely electrons), while the penetrating trigger effectively filters out this soft component by requiring the particle to traverse much more material. This result gives us confidence that a multi-layer coincidence is an effective way to select for a pure muon signal.

Based on this principle, a three-fold coincidence using detectors S1, S2, and S4 was chosen for the main angular distribution measurement. This ``long stack'' configuration was selected for two reasons:

\begin{itemize}
    \item High Purity: By requiring a particle to pass through three layers, including the very bottom one, it provides excellent rejection of the soft component and ensures a very pure muon signal.
    \item Good Angular Resolution: Using the full length of the detector stack (from S1 to S4) creates a telescope with a well-defined geometric acceptance, which is ideal for the angular dependence measurement.
\end{itemize}

\section{Measurement and Analysis of the Muon Angular Distribution}

With the detector performance characterized and the trigger configuration optimized, the main experiment was performed: to measure the rate of cosmic ray muons as a function of the zenith angle $\theta$.

\subsection{Data Collection}

\begin{figure}[h!]
    \centering
    \includegraphics[width=\linewidth]{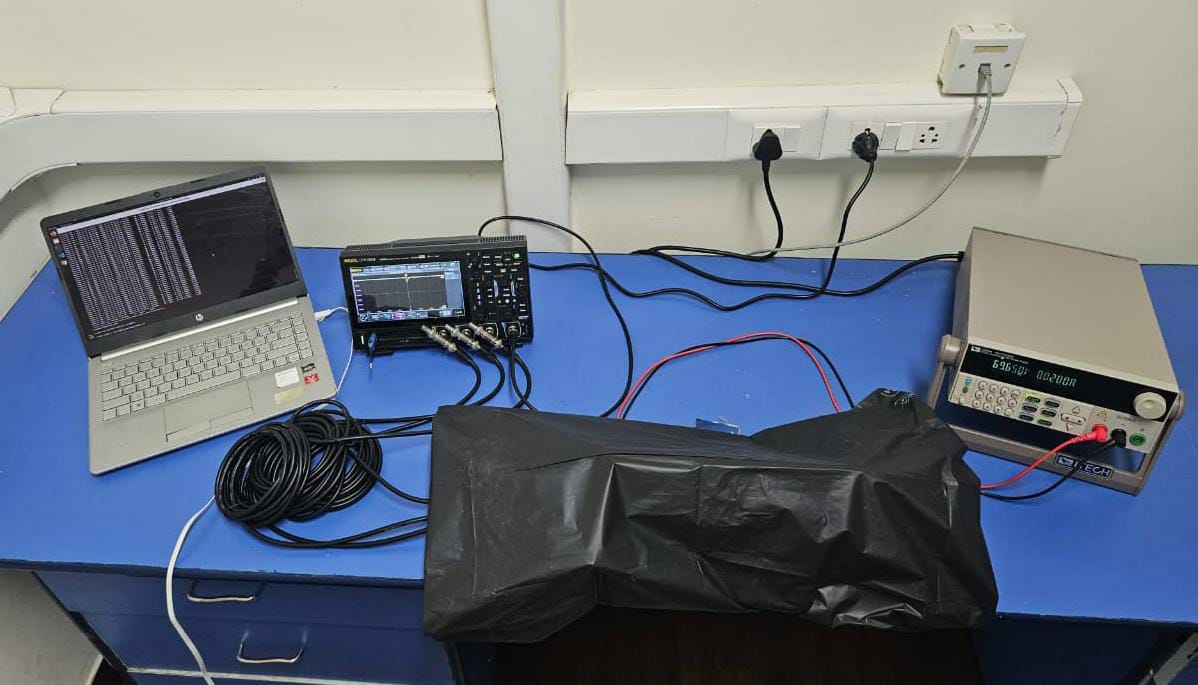}
    
    \caption{The complete experimental setup arranged on the laboratory bench. The system includes the data acquisition laptop, the RIGOL DHO924 oscilloscope, the black-wrapped muon telescope, and the DC power supply.}
    \label{fig:full_setup_photo}
\end{figure}

To establish a high-precision baseline for the vertical muon flux, an extended 12-hour measurement was conducted with the detector positioned at a $0^\circ$ zenith angle. For all other angles, from $10^\circ$ to $90^\circ$ in $10^\circ$ increments, the number of coincidence events was recorded for a total duration of one hour each. The data for all runs was binned into 5-minute intervals to monitor the stability of the count rate. Throughout the data acquisition period, the detector was maintained in complete darkness using a light-proof black sheet (see Fig. \ref{fig:full_setup_photo}) to ensure signal integrity.

\subsection{Data Stability and Final Rate Calculation}

\begin{figure}[h!]
    \centering
    \includegraphics[width=.95\linewidth]{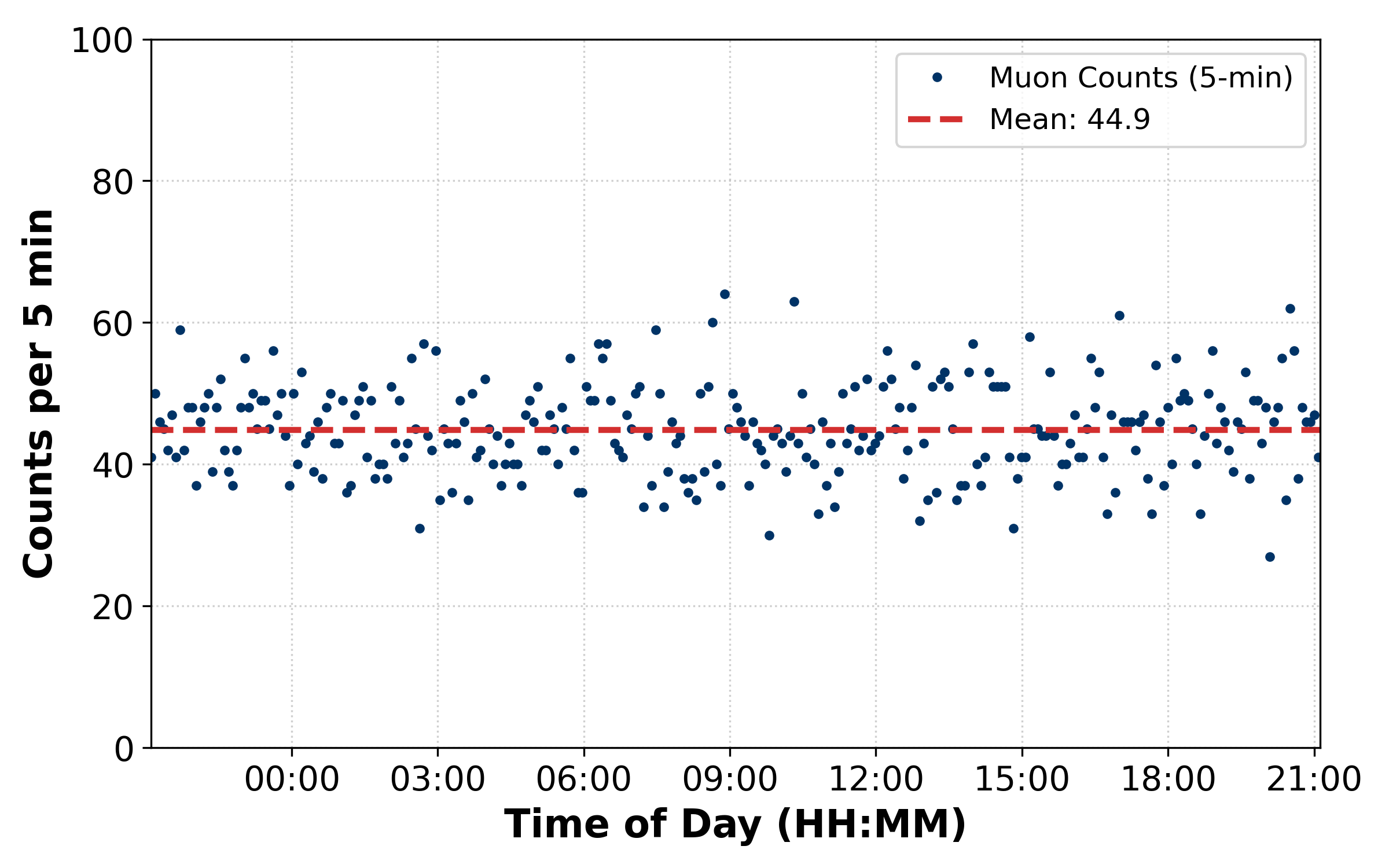}
    \caption{The number of muon events recorded in consecutive 5-minute intervals during the 24-hour measurement at a $0^\circ$ zenith angle. The data points (blue) show the count rate stability over a full day-night cycle, fluctuating randomly around the mean value (red dashed line) without significant drift.}
    \label{fig:stability}
\end{figure}

Before the final analysis, the stability of the data for each run was evaluated. The 1-hour and 24-hour measurements were binned into 5-minute intervals, and the number of counts in each interval was plotted against time to check for drifts or anomalies. A representative stability plot for the long 24-hour measurement at $0^\circ$ is shown in Fig. \ref{fig:stability}.

This stable behavior was found to be consistent across all measured angles, confirming the quality of the collected data.

With the stability of the runs verified, a single average rate and its statistical error were calculated for each angle by dividing the total number of events by the total time. The $0^\circ$ point, benefiting from the much longer run time, has a significantly smaller statistical uncertainty. These final, processed rates are the basis for the main analysis in the following section.

\subsection{Angular Dependence of the Muon Rate}

\begin{figure}[h!]
    \centering
    
    \includegraphics[width=0.95\linewidth]{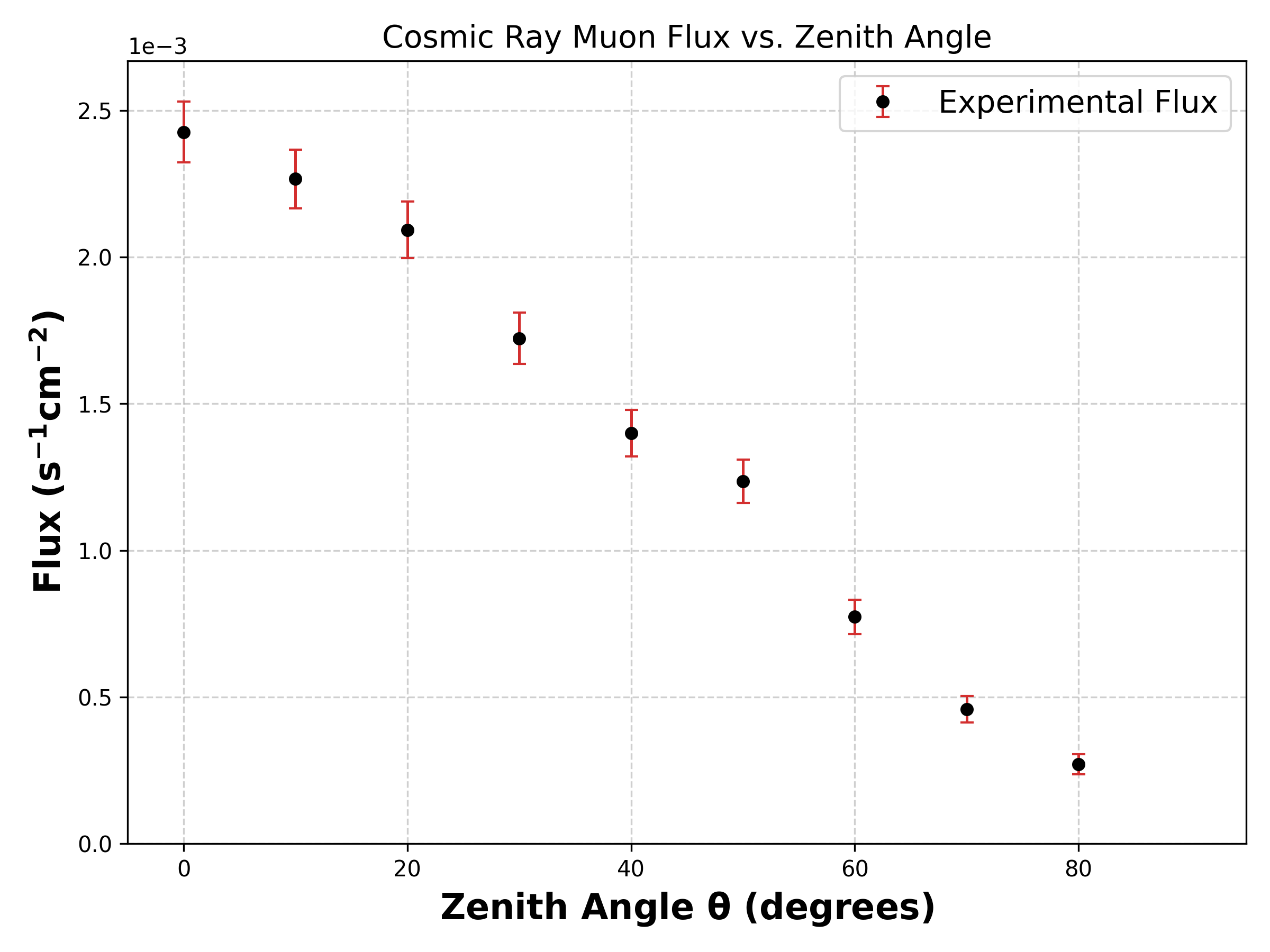}
    
    \caption{The measured cosmic ray muon flux as a function of zenith angle. The flux is calculated by normalizing the measured count rate by the detector area ($62.5~\text{cm}^2$). The data clearly demonstrates a strong directional dependence, with the maximum flux at the vertical ($0^\circ$) and a monotonic decrease towards the horizon. Error bars represent statistical uncertainties ($\pm 1\sigma$).}
    \label{fig:flux}
\end{figure}

The final processed rates for each angle are summarized in Table \ref{tab:final_rates}.
\begin{table}[h!]
    \centering
    \caption{Measured muon rate and calculated flux.}
    \label{tab:final_rates}
    \renewcommand{\arraystretch}{1.2}
    \setlength{\tabcolsep}{4pt} 
    \small 
    \begin{tabular}{cccc} 
        \hline
        \textbf{$\theta$ ($^\circ$)} & \textbf{Rate (Hz)} & \textbf{Err (Hz)} & \textbf{Flux$^*$} \\
        \hline
        0  &  0.152 & 0.0065 & 2.43 \\
        10 &  0.142 & 0.0063 & 2.27 \\
        20 &  0.131 & 0.0060 & 2.09 \\
        30 &  0.108 & 0.0055 & 1.72 \\
        40 &  0.088 & 0.0049 & 1.40 \\
        50 &  0.077 & 0.0046 & 1.24 \\
        60 &  0.048 & 0.0037 & 0.77 \\
        70 &  0.029 & 0.0028 & 0.46 \\
        80 &  0.017 & 0.0022 & 0.27 \\
        \hline
    \end{tabular}
    
    \vspace{1mm}
    \scriptsize 
    
    \textit{*Flux unit: $10^{-3} \cdot s^{-1}cm^{-2}$. Calculated with $A = 62.5~\text{cm}^2$.}
\end{table}
To visualize the relationship between the muon rate and the zenith angle, this data is plotted in Fig. \ref{fig:flux}.

The data in Fig. \ref{fig:flux} clearly demonstrates a strong dependence on the zenith angle. The measured rate is at its maximum at $0^\circ$ and decreases monotonically as the angle increases towards $90^\circ$. This trend is qualitatively consistent with the expected angular distribution \cite{venterea2023analysis}. A detailed quantitative analysis comparing this data to several theoretical models is performed in the next section.

\subsection{Comparative Analysis of Angular Distribution Models}

To perform a quantitative analysis of the data presented in the previous section, the measured rates were fit to several prominent theoretical and phenomenological models from the literature, as discussed in studies such as \cite{venterea2023analysis}. This comparative approach allows for a rigorous evaluation of which model best describes the experimental results. The specific models tested were:
\begin{itemize}
    \item The standard cos-squared approximation.
    \item The general power law model, often referred to as the Pethuraj et al.\ model.
    \item The physically-motivated Shukla and Sankrith model.
    \item The flexible, phenomenological Schwerdt model.
\end{itemize}

\begin{figure}[h!]
    \centering
    \includegraphics[width=0.95\linewidth]{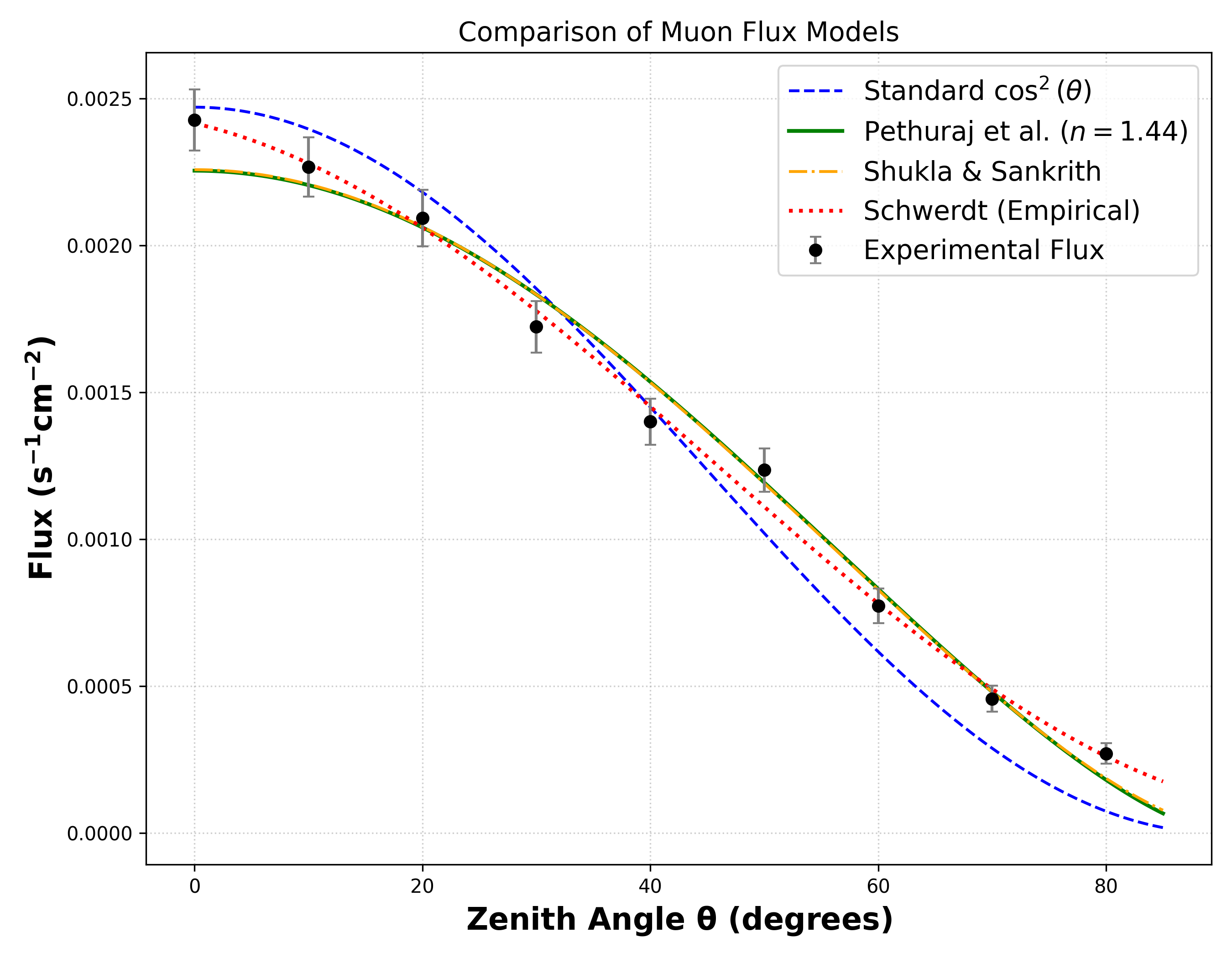}
    
    \caption{Comparison of muon flux models as a function of zenith angle using the data from this experiment. Note that the fit includes angles up to $\theta = 80^\circ$. It can be observed that the Schwerdt model shows excellent agreement with the measured points. The Shukla and Sankrith model and the Pethuraj et al.\ model produce very similar results, leading to a significant overlap in their curves.}
    \label{fig:comparison}
\end{figure}

The best-fit curve for each of these four models is shown overlaid on the experimental data in Fig. \ref{fig:comparison}.

To quantitatively determine which model provides the best description of the data, a Reduced Chi-Squared ($\chi^2/\text{dof}$) test was performed for each fit. The results are summarized in Table \ref{tab:fit_stats}.
\begin{table}[h!]
    \centering
    \caption{Summary of statistical tests for the different fit models.}
    \label{tab:fit_stats}
    \renewcommand{\arraystretch}{1.2}
    \small
    \begin{tabular}{@{} p{2.2cm} c p{4cm} @{}} 
        \hline
        \rule{0pt}{0ex} 
        \textbf{Model} & \textbf{Stats} & \textbf{Key Parameters} \\[1mm] 
        \hline
        
        \rule{0pt}{4ex} 
        Simple $\cos^2$ 
        & \parbox[c]{1.5cm}{\centering $\chi^2_\nu$: 8.36 \\[1ex] $R^2$: 0.9632} 
        & $I_0 = 2.47 \times 10^{-3}$ \\[3mm] 
        \hline
        
        \rule{0pt}{4ex}
        Pethuraj et al
        & \parbox[c]{1.5cm}{\centering $\chi^2_\nu$: 2.29 \\[1ex] $R^2$: 0.9842} 
        & $I_0 = 2.25 \times 10^{-3}$ \newline $n = 1.44 \pm 0.06$ \\[3mm]
        \hline
        
        \rule{0pt}{4ex}
        Shukla 
        & \parbox[c]{1.5cm}{\centering $\chi^2_\nu$: 2.14 \\[1ex] $R^2$: 0.9848} 
        & $n = 2.45$ \\[3mm]
        \hline
        
        \rule{0pt}{4ex}
        Schwerdt 
        & \parbox[c]{1.5cm}{\centering $\chi^2_\nu$: 0.88 \\[1ex] $R^2$: 0.9952} 
        & $a=2.40 \times 10^{-3}$ \newline $b=0.815$ \newline $c=0.135$ \newline $d=5.47 \times 10^{-5}$ \\[3mm]
        \hline
    \end{tabular}
    \vspace{1mm}
    \raggedright
    \footnotesize{*$\chi^2_\nu$ denotes Reduced Chi-Squared ($\chi^2/\text{dof}$).}
\end{table}

Based on the statistical results summarized in Table~\ref{tab:fit_stats}, the Schwerdt model yields the lowest reduced chi-squared value ($\chi^2_\nu = 0.88$), indicating the best overall mathematical agreement with the measured data. However, the model proposed by Pethuraj et al. also provides a very good fit ($\chi^2_\nu = 2.29$) and, importantly, offers a clearer physical interpretation of the angular dependence of the intensity.

Using the Pethuraj et al. model, the vertical intensity is determined to be $I_0 = 2.25 \times 10^{-3}~\text{s}^{-1}\text{cm}^{-2}$, with an angular exponent of $n = 1.44 \pm 0.06$. These parameters characterize the observed angular distribution and are consistent with expectations for atmospheric secondary particles. A more detailed discussion of these results, along with a comparison of the measured exponent $n$ with previously reported values in the literature, is presented in the Discussion section.
\section{Discussion}
Now that the experimental results have been obtained and the relative performance of the different models has been evaluated, this section focuses on interpreting what these results imply physically. In particular, the measured angular dependence is compared with previous experimental studies reported in the literature. In addition, the presence of a constant background component and the possible influence of experimental uncertainties are briefly discussed.

\subsection{The Angular Exponent $n$}

The comparative analysis of the different fitting approaches shows that, while the Schwerdt model provides the best purely mathematical description of the data, the model proposed by Pethuraj et al. offers a physically more meaningful representation of the angular dependence. Consequently, the angular exponent is discussed here based on the Pethuraj et al. model.

From this fit, the angular exponent is determined to be
\begin{equation}
n = 1.44 \pm 0.06 .
\end{equation}

This measured value lies reasonably close to the commonly used theoretical approximation of $n = 2$ \cite{Pethuraj_2017}, which is often employed as a first-order description of the angular distribution. Deviations from this idealized value are expected in real measurements due to factors such as atmospheric attenuation, detector geometry, and local environmental conditions.

To place the present result in a broader context, the measured exponent is compared with values reported in earlier experimental studies. A summary of this comparison is presented in Table~\ref{tab:n_comparison}. The result obtained in this work falls within the range of values reported in the literature, demonstrating consistency with previous measurements while reflecting the specific experimental conditions of the present study.

\begin{table}[h!]
    \centering
    \caption{Comparison of the measured angular exponent $n$ with values reported in the literature.}
    \label{tab:n_comparison}
    \renewcommand{\arraystretch}{1.3}
    \begin{tabular}{lc}
        \hline
        \textbf{Study} & \textbf{$n$ value} \\ 
        \hline
        Our work& \textbf{$1.44 \pm 0.06$} \\
        Judge and Nash~\cite{Judge1965MeasurementsOT} & $1.96 \pm 0.22$ \\
        Crookes and Rastin~\cite{Crookes:1972xd} & $2.16 \pm 0.01$ \\
        Venterea and Ekka~\cite{venterea2023analysis} & $1.39 \pm 0.01$ \\
        \hline
    \end{tabular}
\end{table}

As shown in the Table \ref{tab:n_comparison}, our result of $n = 1.44 \pm 0.06$ falls within the general range of values reported in the literature, though it is notably lower than the theoretical approximation of $n = 2$. The agreement with the study by Venterea and Ekka ($1.39 \pm 0.01$) ~\cite{venterea2023analysis} is particularly strong. This consistency suggests that our detector geometry and local conditions are comparable to those in their experiment, providing high confidence in the validity of our experimental method and analysis.
\subsection{Analysis of the Isotropic Background Component (B or d parameter)}
A key finding from the model fitting was the measurement of a statistically significant, non-zero isotropic background. While we primarily used the Schwerdt model (which includes the offset parameter $d$), this corresponds physically to the background term $B$ in the classic description. From the fit parameters in Table \ref{tab:fit_stats}, this background component was calculated to be:

\begin{equation}
    B = d \times A = (5.47 \times 10^{-5}) \times 62.5 \approx 0.0034~\text{Hz}
\end{equation}

This constant background rate aligns well with the general expectation that at large zenith angles, as the true muon flux significantly diminishes, ambient radiation and other local effects become the dominant signal components~\cite{venterea2023analysis,schwerdt_desy}. This background is likely composed of two main sources: a small, residual rate of accidental coincidences from detector dark counts, and a physical component from cosmic ray side-showers or local ambient radiation that can trigger the detectors from horizontal directions. The inclusion of this $d$ parameter was essential for achieving a good statistical fit, as demonstrated by the significantly lower Reduced Chi-Squared value of the Schwerdt model ($\chi^2_\nu = 0.88$) compared to the simple $\cos^2(\theta)$ model ($\chi^2_\nu = 8.36$).
\subsection{Discussion of Geometric Effects and Sources of Uncertainty}

While the experiment was successful, it is important to discuss potential sources of error. The uncertainties presented on the plots are purely statistical ($\sqrt{N}$). However, several sources of systematic uncertainty could also affect the results.

One potential systematic effect is the geometric acceptance of the telescope. At very small angles ($0^\circ$ to $\sim 20^\circ$), the effective area of the detector can change, which may cause a slight deviation from the pure $\cos^n(\theta)$ law, as was observed in our initial measurements.

Another source of uncertainty is the precision of the angle measurement. Although a digital protractor was used, a small mechanical instability was observed during initial runs, where the detector angle would drift slightly over time. This was mitigated by securely clamping the apparatus for all final measurements, but a residual uncertainty of approximately $\pm 0.5^\circ$ in the angle setting is estimated.

Finally, regarding the model fitting, the Reduced Chi-Squared value for our best-fit model (Schwerdt) was found to be $\chi^2_{\nu} = 0.88$. A value close to 1.0 indicates an excellent agreement between the model and the data. The fact that it is slightly below 1.0 suggests that our statistical error bars were conservatively estimated, covering the observed variations well. This gives us high confidence that the theoretical models successfully describe the physical behavior of the muon flux.
\section{Conclusion}

In this experiment, a three-fold coincidence muon telescope was constructed and characterized using plastic scintillators and MPPC photodetectors. A series of characterization studies, including threshold scans and a penetration experiment, validated the performance of the detector and confirmed its ability to select a pure signal of cosmic ray muons while rejecting background noise.

The primary goal of the experiment was to measure the angular distribution of the muon flux. The measured rate was found to decrease with increasing zenith angle, consistent with theory. A comparative analysis of four prominent models from the literature was performed to quantitatively describe this distribution. While the Schwerdt model provided the lowest chi-squared value, the General $\cos^n(\theta)$ model was utilized to determine the physical angular dependence. From this fit, the angular exponent of the cosmic ray muon flux was measured to be \textbf{$n = 1.44 \pm 0.06$}. This result is in excellent agreement with values reported in the published literature (specifically Venterea and Ekka), though it deviates from the standard theoretical approximation of $n=2$ due to the detector's geometric acceptance and local environmental factors.
\vspace{0mm}

.


%

\appendices

\section{Experimental Operating Parameters}

To facilitate the reproduction of this experiment, the specific electronic settings used for the main data collection are summarized in Table~\ref{tab:op_params}.

\begin{table}[h!]
    \centering
    \caption{Summary of operating parameters used for the angular distribution measurement.}
    \label{tab:op_params}
    \renewcommand{\arraystretch}{1.8}
    \begin{tabular}{l c} 
        \hline
        \textbf{Parameter} & \textbf{Setting / Value} \\
        \hline
        MPPC Bias Voltage ($V_{bias}$) & 69.65 V \\
        Trigger Source & Internal (CH1 \& CH2 \& CH3) \\
        Trigger Slope & Falling Edge (Negative Pulse) \\
        Trigger Threshold & $\approx 3$~p.e. (optimized$^*$) \\
        Coincidence Window ($\tau$) & 200 ns \\
        Logic Operation & \texttt{AND} (S1 $\land$ S2 $\land$ S3) \\
        Run Duration per Angle & 3600 s (1 hour) \\
        \hline
    \end{tabular}
    \vspace{1mm}

    \footnotesize
    \textit{$^*$ \parbox[t]{0.9\linewidth}
    {\centering Note: The threshold was set approximately $100~\mu\text{V}$ below the mean 3~p.e.\ amplitude to maximize detection efficiency while rejecting 2~p.e.\ noise.}}
\end{table}

\section{Data and Code Availability}

The complete experimental dataset and the custom software developed for this study are available in the following repository:

\begin{center}
    \url{https://github.com/msahla99-ops/muon-telescope-project}
\end{center}


\section*{Acknowledgment}
The authors gratefully acknowledge the financial support provided by the Initiation Grant from the Indian Institute of Technology Kanpur. We also thank the Department of Physics for providing the necessary laboratory facilities. We would like to express our sincere gratitude to Son Cao from the \textit{Neutrino Group, IFIRSE, Quy Nhon, Vietnam}, for their valuable contributions and support throughout this project.


%


\printbibliography

@article{janossy1944rate,
  title     = {Rate of n-fold Accidental Coincidences},
  author    = {J\'{a}nossy, L.},
  journal   = {Nature},
  volume    = {153},
  number    = {3875},
  pages     = {165},
  year      = {1944},
  publisher = {Nature Publishing Group},
  doi       = {10.1038/153165a0}
}

@book{leo,
  title     = {Techniques for Nuclear and Particle Physics Experiments: A How-to Approach},
  author    = {Leo, William R},
  year      = {1994},
  publisher = {Springer-Verlag},
  edition   = {2nd},
  address   = {Berlin, Heidelberg}
}

@article{Pethuraj_2017,
  title   = {Measurement of cosmic muon angular distribution and vertical integrated flux by 2 m $\times$ 2 m RPC stack at IICHEP-Madurai},
  author  = {Pethuraj, S. and Datar, V.M. and Majumder, G. and Mondal, N.K. and Ravindran, K.C. and Satyanarayana, B.},
  journal = {Journal of Cosmology and Astroparticle Physics},
  volume  = {2017},
  number  = {09},
  pages   = {021},
  year    = {2017},
  doi     = {10.1088/1475-7516/2017/09/021}
}

@article{Shukla2016EnergyAA,
  title   = {Energy and angular distributions of atmospheric muons at the Earth},
  author  = {Shukla, Prashant and Sankrith, Sundaresh},
  journal = {International Journal of Modern Physics A},
  volume  = {31},
  year    = {2016},
  url     = {https://api.semanticscholar.org/CorpusID:118684516}
}

@article{venterea2023analysis,
  title   = {An Analysis of Muon Flux from Angle Variation of the QuarkNet Cosmic Ray Detector},
  author  = {Venterea, Ricco and Ekka, Urbas},
  journal = {arXiv preprint arXiv:2306.13689},
  year    = {2023}
}

@article{nguyen2023multi,
  title   = {Multi-pixel photon counter for operating a tabletop cosmic ray detector under loosely controlled conditions},
  author  = {Nguyen, Hoang Duy Thanh and others},
  journal = {Dalat University Journal of Science},
  pages   = {16--29},
  year    = {2023}
}

@article{yau2008cosmic,
  title   = {Cosmic ray muon detection using NaI detectors and plastic scintillators},
  author  = {Yau, Chung and Ho, Elton},
  journal = {American Institute of Physics},
  volume  = {3},
  year    = {2008}
}

@phdthesis{le2018cosmic,
  title  = {Cosmic Ray Muon Detection},
  author = {Le Boulicaut, Elise},
  year   = {2018},
  school = {Gustavus Adolphus College},
  type   = {Undergraduate Thesis},
  url    = {https://gustavus.edu/physics/research/studresearch/2018/boulicaut.pdf}
}

@techreport{schwerdt_desy,
  title       = {Zenith Angle Dependence of the Cosmic Muon Rate: Measurement and Analysis with Cosmic@Web},
  author      = {Schwerdt, Carolin},
  institution = {Deutsches Elektronen-Synchrotron (DESY)},
  address     = {Zeuthen, Germany},
  year        = {2020},
  month       = {October},
  type        = {Technical Report},
  url         = {https://icd.desy.de/sites/sites_conferences/site_icd/content/e171345/e171585/toggle_text171586/text171687/Untersuchung_Zenitwinkelabhngigkeit_engl_ICD20.pdf}
}

@book{streetman,
  title     = {Solid State Electronic Devices},
  author    = {Streetman, Ben G. and Banerjee, Sanjay Kumar},
  edition   = {7th},
  year      = {2015},
  publisher = {Pearson Education Limited},
  address   = {Harlow, Essex}
}

@article{blanchard2012measurement,
  title   = {A Measurement of the Angular Distribution of Cosmic Muons},
  author  = {Blanchard, Jennifer},
  journal = {McGill University Physics Dept. Reports},
  volume  = {9},
  pages   = {11},
  year    = {2012}
}

@manual{hamamatsu_tech_note,
  title        = {{MPPC} (Multi-Pixel Photon Counter) Technical Note},
  author       = {{Hamamatsu Photonics K.K.}},
  organization = {Solid State Division},
  number       = {KAPD9008E},
  year         = {2022},
  note         = {Available online},
  url          = {https://www.hamamatsu.com/content/dam/hamamatsu-photonics/sites/documents/99_SALES_LIBRARY/ssd/mppc_kapd9008e.pdf}
}

@article{Judge1965MeasurementsOT,
  title   = {Measurements on the muon flux at various zenith angles},
  author  = {Judge, R. J. R. and Nash, W. F.},
  journal = {Il Nuovo Cimento},
  year    = {1965},
  volume  = {35},
  pages   = {999--1024}
}

@article{Crookes:1972xd,
  author  = {Crookes, J. N. and Rastin, B. C.},
  title   = {An investigation of the absolute intensity of muons at sea-level},
  journal = {Nuclear Physics B},
  volume  = {39},
  pages   = {493--508},
  year    = {1972},
  doi     = {10.1016/0550-3213(72)90384-7}
}

@manual{hamamatsu_s10362,
  title  = {{MPPC} S10362-11 Series Datasheet},
  author = {{Hamamatsu Photonics K.K.}},
  note   = {Cat. No. KAPD1022E05},
  url    = {https://www.phys.hawaii.edu/~idlab/taskAndSchedule/iTOP/SciFi_doco/s10362-11series_kapd1022e05.pdf}
}

\end{document}